\documentclass[acmsmall,screen]{acmart}

\AtBeginDocument{%
  }

\usepackage{subfigure}
\usepackage{colortbl}
\usepackage{soul}  
\usepackage{enumitem}
\usepackage{listings}
\usepackage{graphicx}
\usepackage{fancyhdr}
\usepackage{xcolor}
\usepackage{natbib}
\usepackage{stfloats}

\definecolor{mycolor}{HTML}{9FEE00}
\definecolor{mycolor_green}{HTML}{8AD13D}

\usepackage{wrapfig}

\usepackage{amsmath, amsfonts}
\usepackage[linesnumbered,ruled,vlined]{algorithm2e}
\usepackage{algpseudocode}
\SetKwInOut{Input}{Input}
\SetKwInOut{Output}{Output}

\SetCommentSty{mycommfont}

\DeclareMathOperator*{\argmin}{arg\,min}
\usepackage{multirow}

\definecolor{lavendermist}{rgb}{0.9, 0.9, 0.9}

\acmConference[Conference]{}{2017}{USA}

\theoremstyle{definition}
\newtheorem{definition}{Definition}[]

\begin{document}


\title{Fault Detection for Large Language Models: Benchmark and Enhancement}

\title[Fault Detection for LLMs]{Evaluation and Improvement of Fault Detection for Large Language Models}

\author{Qiang Hu$^*$}
\affiliation{%
  \institution{The University of Tokyo}
  \country{Japan}}

\author{Jin Wen$^*$}\thanks{$^*$ Equal Contribution.}
\affiliation{%
  \institution{University of Luxembourg}
  \country{Luxembourg}}
  
\author{Maxime Cordy}
\affiliation{%
 \institution{University of Luxembourg}
  \country{Luxembourg}}

\author{Yuheng Huang}
\affiliation{%
  \institution{The University of Tokyo}
  \country{Japan}}

\author{Wei Ma}
\affiliation{%
  \institution{Nanyang Technological University}
  \country{Singapore}}

\author{Xiaofei Xie}
\affiliation{%
 \institution{Singapore Management University}
  \country{Singapore}}

\author{Lei Ma}
\affiliation{%
  \institution{The University of Tokyo \& University of Alberta}
  \country{Japan \& Canada}}

\begin{abstract}
Large language models (LLMs) have recently achieved significant success across various application domains, garnering substantial attention from different communities.
Unfortunately, even for the best LLM,  many \textit{faults} still exist that LLM cannot properly predict. Such faults will harm the usability of LLMs in general and could introduce safety issues in reliability-critical systems such as autonomous driving systems. How to quickly reveal these faults in real-world datasets that LLM could face is important, but challenging. The major reason is that the ground truth is necessary but the data labeling process is heavy considering the time and human effort. To handle this problem, in the conventional deep learning testing field, test selection methods have been proposed for efficiently evaluating deep learning models by prioritizing faults. However, despite their importance, the usefulness of these methods on LLMs is unclear, and lack of exploration. In this paper, we conduct the first empirical study to investigate the effectiveness of existing fault detection methods for LLMs. Experimental results on four different tasks~(including both code tasks and natural language processing tasks) and four LLMs~(e.g., LLaMA3 and GPT4) demonstrated that simple methods such as Margin perform well on LLMs but there is still a big room for improvement. Based on the study, we further propose \textbf{MuCS}, a prompt \textbf{Mu}tation-based prediction \textbf{C}onfidence \textbf{S}moothing framework to boost the fault detection capability of existing methods. Concretely, multiple prompt mutation techniques have been proposed to help collect more diverse outputs for confidence smoothing. The results show that our proposed framework significantly enhances existing methods with the improvement of test relative coverage by up to 70.53\%.

\end{abstract}

\maketitle


\section{Introduction}
\label{sec:intro}
In recent years, the potential of large language models~(LLMs) has brought hope to general artificial intelligence~(AGI). LLMs achieved promising and even the best results compared to other conventional methods in many areas, such as question answering~\cite{zhou2022large}, sentiment analysis~\cite{zhang2023sentiment}, and code understanding~\cite{ma2023scope} that make LLMs become the first choice of deep learning models in such tasks. Besides the most famous application ChatGPT~\footnote{OpenAI, ChatGPT, \url{https://chat.openai.com/}}, multiple LLMs have been proposed, e.g., LLaMA~\cite{touvron2023llama}, Alpaca~\cite{taori2023alpaca}, and Cerebras-GPT~\cite{dey2023cerebras}. More interestingly, focusing on software engineering~(SE) tasks, a series of code-related LLMs with good programming ability have been released to help the software development, such as StarCoder~\cite{li2023starcoder} and CodeLLaMA~\cite{roziere2023code}. Learning to use LLMs is already a basic skill for us in daily life.

Despite the success, the same as other types of deep learning models, LLMs also suffer from different problems that limit their practical usage. For example, LLMs sometimes generate unrelated outputs to the inputs or incorrect results misaligned with established world knowledge, which is known as the hallucination problem~\cite{zhang2023siren}. Besides, recent work~\cite{dong2023robust} showed that LLMs can be easily fooled by adversarial attacks and are not robust. Furthermore, considering SE tasks, existing code-related LLMs cannot capture equivalent semantics when the given natural language prompt expresses the same meaning in different languages~\cite{peng2024xl}. Those limitations remind us that LLMs are not always trustworthy, and it is necessary to carefully evaluate LLMs and quickly reveal the unreliable outputs of LLMs in practical usage.

The basic way to evaluate LLMs and reveal their weakness is to prepare test data as diverse as possible to assess the performance of LLMs accordingly. However, preparing such test data is heavy work due to the complicated process of data collection, data cleaning, and data labeling. Especially, data labeling requires human effort with domain knowledge which is the most challenging part. Moreover, some LLMs are closed-source which requires funding expenses, e.g., it tasks 0.03\$ per 1,000 tokens of inputs for GPT4 using OpenAI API. Therefore, it is almost impossible to thoroughly evaluate LLMs on the target downstream task for people who have limited budgets. As a result, how to efficiently test LLMs and filter the mispredicted data~(unreliable outputs) of LLMs with less effort becomes an urgent problem.

To tackle the data labeling issue mentioned above, in the field of deep learning testing, multiple fault detection methods that quickly identify fault data~(data the model has incorrect prediction) in the test set without using the labeling information~\cite{10.1145/3643678} have been proposed. Based on the information required, existing methods can be divided into learning-based methods and output-based methods. Those methods have been widely studied in the classical deep learning models and have proven to be effective in saving the labeling budget during testing. However, such studied classical deep learning models are constructively different from LLMs, e.g., LLMs are pre-trained using a large amount of pre-training data with diverse tasks which makes their prediction confidence~(most fault detection methods rely on) unclear given a specific downstream task. Therefore, as no studies have explored the effectiveness of fault detection methods for LLMs, it is unknown whether we can directly employ existing methods to evaluate LLMs and save our effort or not. 

To bridge this gap, in this paper, we conduct the first study to investigate the effectiveness of existing fault detection methods for LLMs. The major challenge in the study is that most fault detection methods are proposed for classification tasks and use the output probabilities for data prioritization, but LLMs normally output a sequence of tokens without such a clear output probability for the given task. To solve this, we design the prompt to guide LLMs to produce the output with their confidence. For datasets that have more than two classes, we add examples~(few-shot) in each class in the prompt as guidance to ask LLMs to predict new data. In total, our study covers nine fault detection methods including both learning-based methods, e.g., TestRank~\cite{NEURIPS2021_ae785101}, and output-based methods, e.g., ATS~\cite{10.1145/3510003.3510232}. Based on the experiment results collected from four datasets~(e.g., sentiment analysis and code clone detection) and four LLMs~(e.g., LLaMA3 and GPT4). We found that 1) LLMs are not well-calibrated and overconfident in clone detection, problem classification, and news classification tasks, 2) simple methods such as Margin perform better compared to others, and 3) however, existing fault detection methods perform relatively pool on LLMs compared their performance on classical deep learning models. There is a need to propose methods to further enhance existing fault detection methods.

To do so, inspired by~\cite{10.1109/ICSE48619.2023.00110, 10.1109/ICSE43902.2021.00046} which utilize mutation testing techniques to help test deep learning models, we introduce \textbf{MuCS}, a prompt \textbf{Mu}tation-based prediction \textbf{C}onfidence \textbf{S}moothing framework to enhance existing fault detection methods. Specifically, we mutate the prompt by transformation methods to generate a set of prompt mutants first. Here, both text augmentation methods and code refactoring methods have been considered in MuCS. Then, we collect the output probabilities of all mutants and compute the average prediction confidence. Finally, we use the averaged confidence to perform fault detection using current fault detection methods. We compare the effectiveness of fault detection methods with and without our confidence smoothing method and found that using MuCS significantly enhances the performance of existing methods by up to 70.53\%.

To summarize, the main contributions of this paper are:

\begin{itemize}[leftmargin=*]
    \item  This is the first study that explores the effectiveness of fault detection methods on LLMs. We found that 1) LLMs are overconfident in clone detection, problem classification, and news classification tasks; 2) the prediction confidence of LLMs is concentrated in a few intervals; and 3) simple methods perform relatively better than others.
        
    \item We propose MuCS, a prompt mutation-based method to smooth the confidence of LLMs and enhance the effectiveness of existing fault detection methods.
 
\end{itemize}

The rest of this paper is organized as follows. Section~\ref{sec:related_work} reviews the related works. Section~\ref{sec:design} presents the design of our empirical study and Section~\ref{sec: empirical_results} summarizes the empirical results. Section~\ref{sec:mutation} introduces our proposed mutation-based confidence smoothing solution and Section~\ref{sec:mutation_evaluation} presents its related evaluation. Section~\ref{sec:discussion} discusses the limitation of this work and Section~\ref{sec:conclusion} concludes this work.

\section{Related Work}
\label{sec:related_work}

We review the related works from the perspectives of fault detection for deep neural networks and large language model testing.

\subsection{Fault Detection for Deep Neural Networks}

To save the labeling budget, multiple test selection methods~\cite{al2023deepabstraction++, zheng2023certpri, zhu2023openmix, yin2023dynamic} have been proposed. A recent survey~\cite{10.1145/3643678} reviewed test selection methods and divided them into fault detection methods, sampling-based model retraining methods, model selection methods, and performance estimation methods. We focus on the fault detection methods. 

The early method DeepGini~\cite{10.1145/3395363.3397357} has been proposed by Feng~\emph{et al.} to reveal inputs that are more likely been mispredicted
by the model. DeepGini~\cite{10.1145/3395363.3397357} defined a new Gini score to measure the uncertainty of deep learning models on the inputs. Later on, inspired by the mutation testing in the conventional SE field, Wang~\emph{et al.}~\cite{10.1109/ICSE43902.2021.00046} proposed to mutate both input samples and deep learning models and identify faults by computing the killing score. Their evaluation on more than 20 tasks demonstrated that mutation testing is useful for helping find \textit{bugs} in deep learning models. Furthermore, Gao \emph{et al.}~\cite{10.1145/3510003.3510232} proposed an adaptive test selection~(ATS) method. ATS not only considers detecting faults but also tries to find diverse faults~(diverse here means faults are from different categories) in deep learning models. More recently, Bao~\emph{et al.}~\cite{10.1145/3597926.3598073} first empirically proved that simple methods~(i.e., output probability-based methods) perform well on fault detection. Then, they proposed a new method to compute the uncertainty scores by averaging the basic score of the input and scores from the neighbors of this input. Besides, Ma~\emph{et al.}~\cite{ma2021test} conducted a comprehensive empirical study to explore the potential of test selection methods and active learning methods on fault detection. They found that MaxP is the best in terms of the correlation between the MaxP score and the misclassification. Besides the SE community, researchers from the ML community also contributed to this field. Dan~\emph{et al.}~\cite{hendrycks2016baseline} built the first benchmark for detecting misclassified data and found that just using the confidence score is already a strong baseline for this task. Li~\emph{et al.}~\cite{NEURIPS2021_ae785101} proposed TestRank which uses graph neural networks to learn the difference between benign inputs and faults for fault detection.

Different from these works, our work is the first to explore the effectiveness of fault detection methods for large language models. We found that existing methods cannot detect faults in LLMs well and proposed the use of prompt mutation to further enhance these methods.

\subsection{Large Language Model Testing}

Large language models have been the most used deep learning models for most of the tasks. Comprehensively testing or evaluating the performance of LLMs has become one of the most important directions in almost all communities.

Chang \emph{et al.} surveyed works that focus on evaluating LLMs~\cite{chang2023survey}. They classified the evaluation objectives into seven categories, natural language processing, robustness/ethics/biases/trustworthiness, social science, natural science \& engineering, medical applications, agent applications, and other applications. Evaluating LLMs for code which we are more interested in lies in the category of natural science \& engineering. Xu \emph{et al.}~\cite{xu2022systematic} conducted a comprehensive study to evaluate four series of LLMs on the code generation task using the HumanEval benchmark. They found that the performance of LLMs is affected by the parameter size of models and the training time. Liu \emph{et al.}~\cite{liu2024your} propose a new code synthesis evaluation framework EvalPlus to measure the correctness of code generated by LLMs. Based on this framework, they found that the existing benchmark HumanEval is not rigorous enough. The performance~(pass\@k) of LLMs drops a lot when using the new benchmark. Besides, Ma \emph{et al.}~\cite{ma2023scope} evaluated the capability of LLMs to understand code syntax and semantics. The experimental results show that LLMs can understand the syntax structure of code, and perform code static analysis, but cannot approximate code dynamic behavior of code.

More recently, Yuan \emph{et al.}~\cite{yuan2023evaluating} compared LLMs with conventional code models that are fine-tuned for specific code tasks. They found that in the zero-shot setting, instruction-tuned LLMs have competitive performance to fine-tuned code models. In the one-shot setting, the guidance by one-shot example sometimes harms the performance of LLMs. Du \emph{et al.}~\cite{du2024evaluating}  argued that existing works that evaluate the effectiveness of LLMs for code mainly focus on simple code tasks~(function-level code synthesis) and built the first class-level code generation benchmark for the LLM evaluation. Based on their benchmark, they found that existing LLMs have significantly lower performance on class-level code generation tasks compared to method-level code generation tasks. There is still a big room to improve the ability of LLMs for code generation.

Existing LLM testing works mainly focus on constructing new and challenging datasets (by manual data collection and adversarial data generation, etc.) to validate the capability of LLMs, which can be seen as test augmentation techniques. In contrast, our work targets test selection from the perspective of software testing for LLM testing.

\section{Empirical Study}
\label{sec:design}

Our study focuses on the problem of fault detection of LLMs. We consider both learning-based fault detection methods and output-probability-based methods in the literature. Based on the previous definition~\cite{10.1145/3643678}, given a LLM $M$, an unlabeled test set $X_{test}$, and a labeling budget $budget$, fault detection is to select a subset $\widetilde{X}$ of $X_{test}$ such that $\widetilde{X}_{budget} = \argmin_{\{X_i\subseteq X_{test} \wedge \left|X_i\right| = budget\}} \varrho (M, X_i, Y_i)$ where $Y_i$ are the labels corresponding to $X_i$, and $\varrho\left(\cdot\right)$ is the performance measurement function. We follow the guidance of previous works~\cite{10.1145/3643678, ma2021test, hu2023evaluating} to select nine fault detection methods in our study. We exclude neuron coverage-based methods and surprise adequacy methods in our study since the parameter size of LLMs is too big~(\textgreater 7B) and the coverage and adequacy score are difficult to compute. 

\subsection{Fault Detection Methods}

In total, we collect nine fault detection methods. Let $\mathcal{X}$ be a set of test inputs and $\mathcal{Y}$ be the corresponding label set, $x\in\mathcal{X}$ and $y\in\mathcal{Y}$ denote a given test sample and its true label, respectively.  Let $p_{y_i}\left(x\right)$ be the likelihood of $x$ belonging to class $y_i\in\mathcal{Y}$ produced by the model $\mathcal{M}$.

\begin{itemize}[leftmargin=*]

\item \textbf{Random Selection} is the basic selection criteria that test the model using randomly selected a subset of test data.

\item \textbf{Max Probability~(MaxP)} prioritizes data based on the maximum output probability, expressed as $\mathsf{Max}\left(p_{y} \mid y \in \mathcal{Y}\right)$. Data samples with lower MaxP scores are more likely to be faults.

\item \textbf{DeepGini} defines a Gini score calculated by $Gini\left(x\right)=1-\sum_{y_i\in\mathcal{Y}}{p_{y_i}}^2\left(x\right)$ to measure the uncertainty of data. The data with higher Gini scores are treated as faults.

\item \textbf{Entropy} selects data with the minimum Shannon entropy calculated using output probabilities. 

\item \textbf{Margin} measures the uncertainty of data based on the difference between the top-1 and top-2 output probabilities. A smaller difference indicates that the model is more difficult to distinguish the data between the two classes and the data should be considered faults.

\item \textbf{Multiple-Boundary Clustering and Prioritization~(MCP)} contains two steps. First, it divides the data space into different areas based on the decision boundary. Here, the decision boundary is approximated by the top-2 prediction probabilities. For example, give an input $x$ and its top-2 output probabilities are $p_{y_1}$ and $p_{y_3}$. Then this input is near the decision boundary $(1, 3)$ and on the side of $1$. MCP then selects data from different data spaces based on the score $\frac{p_{y_{f}}}{p_{y_{s}}}$, where $y_f$ and $y_s$ are the top-1 and top-2 probabilities. 

\item \textbf{Nearest Neighbor Smoothing (NNS)} first finds the neighbors and then smooths the output probabilities of each input using the outputs of neighbors. Finally, uncertainty-based fault detection methods such as DeepGini are employed to select the faults. In NNS, neighbors are searched by computing the distance between each extracted representation of the input.  NNS requires the intermediate outputs of inputs to conduct data selection and, thus, cannot be used for closed-source LLMs such as GPT3.5.

\item \textbf{Adaptive Test Selection~(ATS)} first projects the top-3 maximum output probabilities to the space plane. After that, it computes the coverage of each input on the plane. Then, the coverage score is used to identify if the input is a fault or not based on its difference from the coverage of the whole test set.  Note that ATS can only be used for classification tasks that have greater than three categories.

\item \textbf{TestRank} initially extracts two features from the input data: 1) the output from the logits layer as intrinsic attributes, and 2) the graph information, which includes the cosine distance to other data points and the label of the data. Subsequently, a graph neural network (GNN) model is employed to assimilate the graph information and forecast the contextual attributes of the data. Finally, the contextual attributes and intrinsic attributes are amalgamated and inputted into a binary classification model to acquire proficiency in identifying failures. Comparable to the NNS method, TestRank also necessitates intermediate outputs for data selection and is not applicable to closed-source LLMs.

\end{itemize}

Except for the random selection and TestRank, all the other methods are purely based on the output probability. For the TestRank, the output probability is also a part of the information used for the detection. Thus, the prediction confidence of LLMs is important for fault detection.



\subsection{Datasets and Models}
\label{sec:datasets}

\begin{table}[]
\caption{Datasets and models used in our study.}
\centering
\label{tab:datasets}
\resizebox{0.75\textwidth}{!}
{
\begin{tabular}{lcccc}
 \bottomrule
\textbf{Task} & \multicolumn{1}{l}{\textbf{Class Number}} & \textbf{Model} & \multicolumn{1}{l}{\textbf{Test Size}} & \textbf{Accuracy} \\ \bottomrule
\multirow{3}{*}{\textbf{Clone Detection (CD)}} & \multirow{3}{*}{2} & DeepSeekCoder & 1000 & 69.4\% \\
 &  & GPT3.5 & 200 & 75.5\% \\
 &  & GPT4 & 200 & 77.0\% \\ \bottomrule
\multirow{3}{*}{\textbf{Problem Classification (PC)}} & \multirow{3}{*}{5} & DeepSeekCoder & 1000 & 42.6\% \\
 &  & GPT3.5 & 200 & 36.0\% \\
 &  & GPT4 & 200 & 100.0\% \\ \bottomrule
\multirow{3}{*}{\textbf{Sentiment Analysis (Sentiment)}} & \multirow{3}{*}{3} & LLaMA3 & 1000 & 68.8\% \\
 &  & GPT3.5 & 150 & 82.0\% \\
 &  & GPT4 & 150 & 90.0\% \\ \bottomrule
\multirow{3}{*}{\textbf{TagMyNews (TMN)}} & \multirow{3}{*}{7} & LLaMA3 & 1400 & 46.7\% \\
 &  & GPT3.5 & 200 & 60.0\% \\
 &  & GPT4 & 200 & 82.0\% \\ \bottomrule
\end{tabular}
}
\end{table}

We consider four datasets including both natural language classification tasks and programming language classification tasks, and four types of LLMs covering both open-sourced models and closed-source models. Due to the high cost of assessing close-source LLMs, we prepare relatively more test data for open-source LLMs than close-source LLMs.

\textbf{Clone Detection~(CD)}: Utilizing the BigCloneBench~\cite{bigclonebench} dataset, we randomly select 1000 and 200 items to investigate the code clone detection ability of open-source LLMs and closed-source LLMs, respectively. Models are evaluated on their ability to predict the presence of code clones (a probability score from 0 to 1, where 0 indicates no clone and 1 indicates a clone) between code snippet pairs, a critical task for enhancing software maintenance and development practices. Note that, some fault detection methods that require more than two classes such as ATS cannot apply to clone detection tasks.

\textbf{Problem Classification~(PC)}: Based on the Java250~\cite{java250} dataset from the CodeNet Project, problem classification focuses on the classification of programming challenges. 1000 samples and 200 samples across 5 types of problems are randomly chosen for evaluating open-source and closed-source LLMs. This task measures the model's proficiency in categorizing coding problems using a singular example for each category, requiring output as a probability distribution across classes that sum to 1. We employ the one-shot setting~(one-shot examples are randomly selected from the original dataset) for this task due to the poor performance of zero-shot evaluation~(nearly random guess).

\textbf{Sentiment Analysis~(Sentiment)}: For this task, 1000 samples and 150 samples are extracted from the Sentiment Analysis of IMDB Movie Reviews dataset~\footnote{Autolabel, \url{https://github.com/refuel-ai/autolabel}} to evaluate open-source and closed-source LLMs, respectively. The challenge for models is to predict the sentiment of movie reviews, negative, neutral, and positive, ranging from 0 to 1, evaluating the capacity of models for natural language understanding and emotional tone assessment.

\textbf{TagMyNews~(TMN)}: Using the TagMyNews Dataset, we randomly picked 1400 and 200 articles with seven categories, which are sourced from RSS feeds of popular newspapers~\cite{tag_my_news} for open-source and closed-source LLMs, respectively. This task aims to assess the efficacy of the model in accurately categorizing news articles based on their textual content. Due to the same reason as the problem classification task, we employ the one-shot setting for this task.

\textbf{LLaMA3 and DeepSeekCoder.} 
LLaMA3~\cite{touvron2023llama} is known for its auto-regressive architecture aimed at chat applications. The parameters of the series of LLaMA3 models range from 8 billion to 70 billion and are learned from publicly available online data amounting to 2 trillion tokens. Besides, LLaMA3 models have been fine-tuned with a combination of supervised fine-tuning (SFT) and reinforcement learning with human feedback~(RLHF), focusing on aligning the models to human preferences for helpfulness and safety. 
DeepSeeker-Coder consists of a series of code language models, which achieve competitive performance on different code-related tasks~\cite{deepseek24}.
DeepSeek-Coder not only beats most open-sourced LLMs but also surpasses existing advancing close-source LLMs like Codex and GPT-3.5.
We employ the instruct version~(meta-llama/Meta-Llama-3-8B-Instruct and deepseek-ai/DeepSeek-Coder-V2-Lite-Instruct) available on the public Hugging Face platform for our experiments.

\begin{wrapfigure}{r}{0.45\textwidth}
 \centering
 \includegraphics[height=0.5\textwidth]{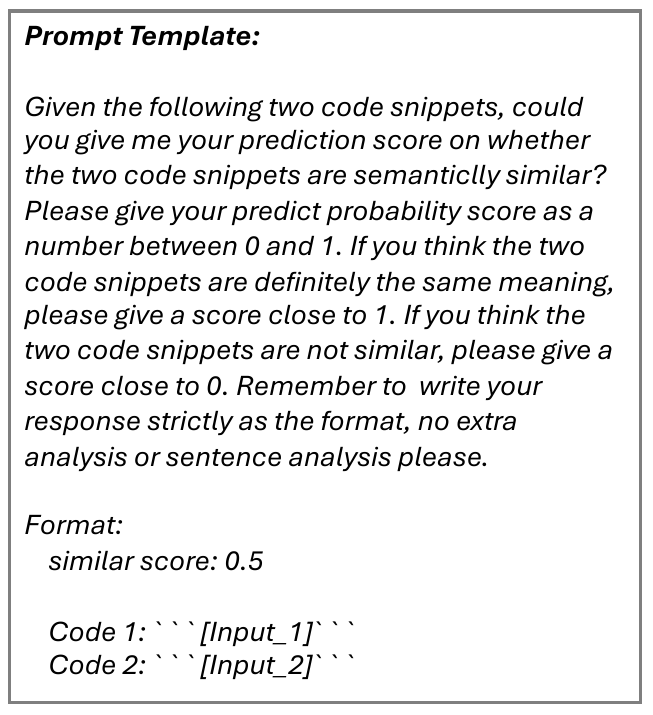}
 \caption{Prompt template of clone detection.}\label{fig:prompt}
 \vspace{-5mm}
\end{wrapfigure}

\textbf{GPT3.5 and GPT4}~\cite{achiam2023gpt}.
GPT3.5 is a fined-tuned version of GPT3 developed in January 2022. It is built on Transformer architecture while emphasizing decoder-only mechanisms. GPT4, which is known as the state-of-the-art LLM, has a substantial leap in capability and complexity compared to GPT-3.5. Both GPT3.5 and GPT4 are closed-source LLMs, we employ \textit{gpt-3.5-turbo-0125} and \textit{gpt-4-0125-preview} models in our experiments through API access, specifically within text chat application contexts.

Table~\ref{tab:datasets} presents the details of our used datasets and models. We found that in some cases~(marked by red), the performance of LLMs is too bad~(close to random guess) and too good~(with perfect accuracy). For these cases, we only evaluate their prediction confidence~(RQ1) and do not conduct fault detection~(RQ2) on them.

\subsection{Prompt Design}


The quality of input prompts highly affects the output of LLMs. To obtain the output probabilities of LLMs given the input, we add guidance in the prompt to ask LLMs to produce their prediction confidence of each category. Figure~\ref{fig:prompt} presents the template example of the code clone detection task. For multiple class classification tasks such as problem classification, we use one-shot examples to lead LLMs to produce more accurate prediction confidence.

\subsection{Research Question}

Our study aims to explore the effectiveness of existing fault detection methods on LLMs. Concretely, we answer the following two research questions: 

\begin{itemize}[leftmargin=*]
    \item \textbf{RQ1: How confident are LLMs in their prediction?} Most~(seven out of nine) of the fault detection methods are purely output uncertainty-based. Therefore, checking the prediction confidence~(an important indicator to measure the output uncertainty) produced by LLMs before running fault detection methods is necessary. 

    \item  \textbf{RQ2: How effective are fault detection methods for LLMs?}  In this research question, we check if the existing methods demonstrated to be useful in classical deep learning models can perform well on LLMs. 
 
\end{itemize}

\subsection{Evaluation Metrics.}

In RQ1, we follow the previous work~\cite{guo2017calibration} to evaluate the confidence of LLMs in their prediction using metrics \textit{Prediction Confidence~(Confidence)} and \textit{Expected Calibration Error~(ECE)}. Roughly speaking, \textit{Confidence} is computed by probabilities associated with the predicted label. \textit{ECE} first splits the range of the prediction score into $M$ equally-spaced intervals (denoted as $I_m$, $m\in \{1,\ldots, M\}$) and then computes the weighted average of the bins’ accuracy/confidence. A lower ECE score indicates that the model is better calibrated, i.e., the prediction probabilities can better represent the true correctness of models.

\begin{definition}[\textbf{Prediction Confidence~(Confidence)}]
\label{def:confidence}

$$Confidence(x) = max(\{p_{y_i}(x) |0<i\leq N\})$$

where $N$ is the number of classes.

\end{definition}

\begin{definition}[\textbf{Expected Calibration Error~(ECE)}] Given a test set $X$, let $B_m \subseteq X$ be the inputs whose prediction confidence falls into the $m^{th}$ interval $I_m=(\frac{m-1}{M}, \frac{m}{M}]$, $Acc(B_m)$ be the accuracy of the inputs $B_m$, and $Avg\_Conf(B_m)=\frac{1}{|B_m|}\sum_{x\in B_m}Confidence(x)$ be the average confidence within $B_m$, then the expected calibration error on $X$ is defined as:

    $$ECE(X) = \sum\limits_{m = 1}^{M}\frac{|B_{m}|}{|X|}\left|Acc(B_{m}) - Avg\_Conf(B_{m})\right|$$
\end{definition}

In RQ2, we follow the previous work~\cite{NEURIPS2021_ae785101} and evaluate the effectiveness of fault detection methods using the metric of Test
Relative Coverage~(TRC) which is calculated by the percentage of the number of faults divided by the labeling budget or the total number of faults whichever is minimal.

\begin{definition}[\textbf{Test
Relative Coverage~(TRC)}] Given a test set $X$ and a labeling budget $budget$, the test relative coverage on $X$ is defined as:

$$TRC(X)=\frac{|\widetilde{X}_{faults}|}{Min\left(budget, |X_{faults}|\right)}$$

\noindent where $\widetilde{X}_{faults}$ represents the faults within the selected test set $\widetilde{X}_{budget}$ and $X_{faults}$ represents the faults in the whole test set $X$. A higher TRC value indicates the better performance of the fault detection method.
\end{definition}

\subsection{Configurations}

\textit{Configuration of fault detection methods.} NNS and TestRank need to compute the distance between data samples and require the representation of the samples. In this work, we directly use the input embeddings extracted from LLMs as the representation. Besides, TestRank is a learning-based method where training data is required for fault detection. For the clone detection, problem classification, and TagMyNews tasks, we randomly select 200 data from the original datasets as the training data. For the sentiment analysis dataset, as the original dataset only contains 200 data samples and 150 of them are used as test data, we collect the remaining 50 samples as the training data. For the data labeling process, we set the labeling budgets from 10\% to 90\% with an interval of 10\%. 

\textit{Configuration of LLMs.} In our experiments, we set \textit{top\_p=0.95} for open-sourced LLaMA3 and DeepSeek-Coder, and \textit{top\_p=1} for closure GPT3.5 and GPT4 which are the default settings used by LLMs. \textit{top\_p} controls the accumulation of token probabilities to select the next word for all LLMs. We set the window size of the token context as \textit{max\_length= 10k} for LLaMA3 and DeepSeek-Coder to match our test data's token length. The window sizes of GPT3.5 and GPT4 are set as 16385 and 128000 respectively, using the default settings.

\subsection{Implementation and Environment}

For the fault detection methods, we reuse the official implementations provided by the original papers and modify them to fit our tasks. For the LLaMA and CodeLLaMA models, we use the resources provided by Hugging Face\footnote{Hugging Face, \url{https://huggingface.co/}} in this work. For GPT3.5 and GPT4 models, we use the OpenAI's API to access the models and collect the results. In total, it costs 267 US dollars to run GPT-related experiments. We conduct all LLaMA-related experiments on a high-performance computer cluster.  Each cluster node runs a 2.6 GHz Intel Xeon Gold 6132 CPU with an NVIDIA Tesla V100 16G SXM2 GPU.

\section{Empirical Results}
\label{sec: empirical_results}

\subsection{RQ1: Prediction Confidence}
\label{sec:rq1}

\begin{figure*}[h]
    \centering
    \subfigure[CD, DeepSeekCoder, \textbf{ECE: 0.9421}]{
    \includegraphics[scale=0.26]{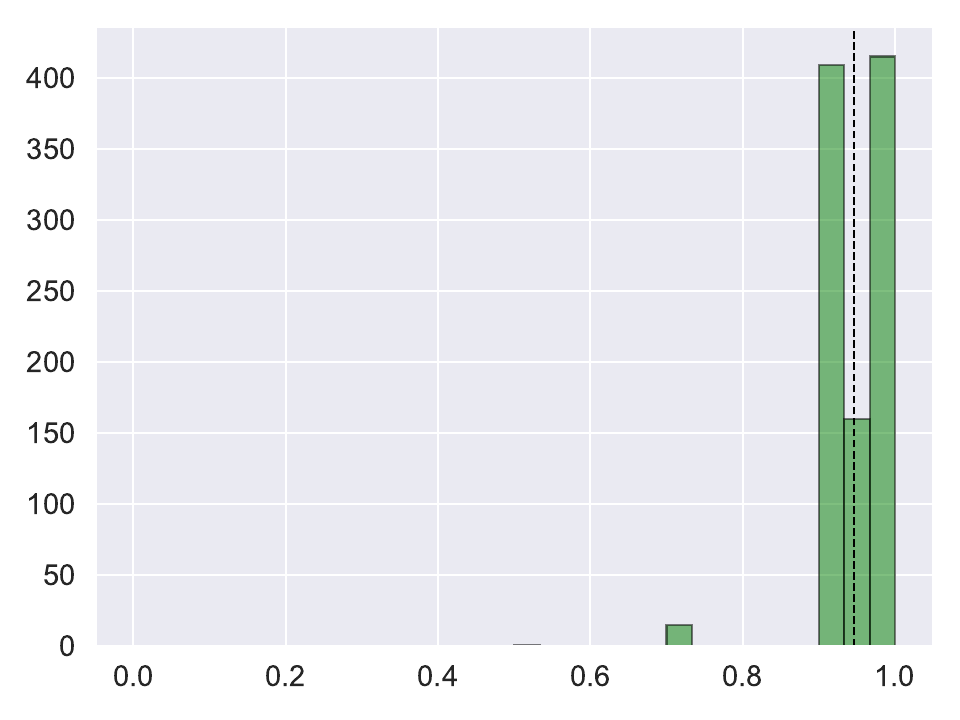}%
    }
    \subfigure[PC, DeepSeekCoder, \textbf{ECE: 0.3110}]{
    \label{fig:pc_llama}
    \includegraphics[scale=0.26]{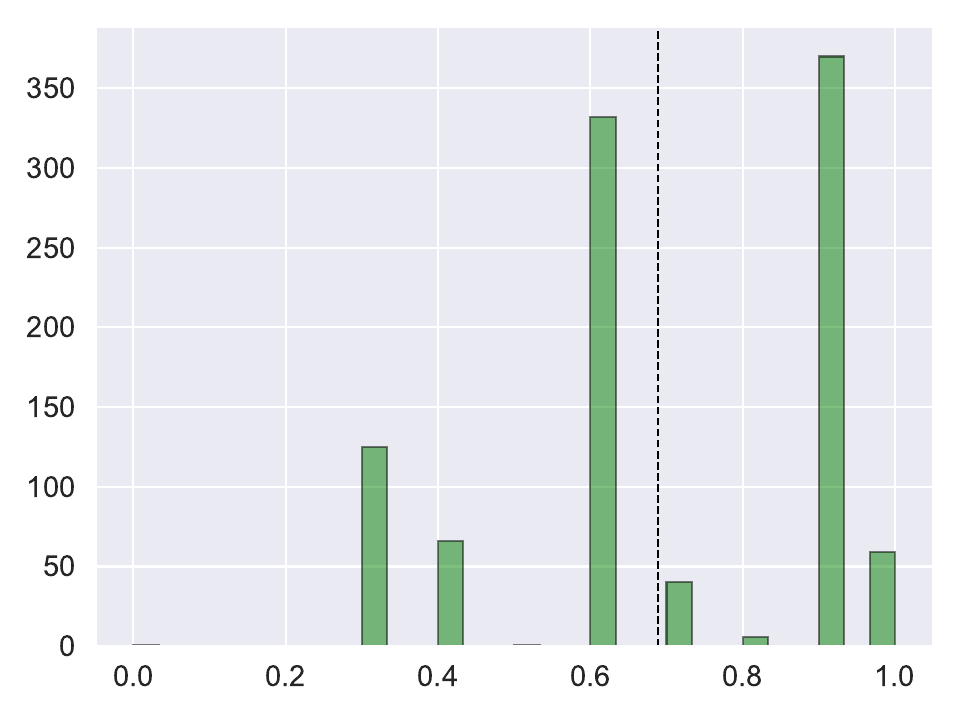}%
    }
    \subfigure[TMN, LLaMA3, \textbf{ECE: 0.1711}]{
    \label{fig:tmn_llama}
    \includegraphics[scale=0.26]{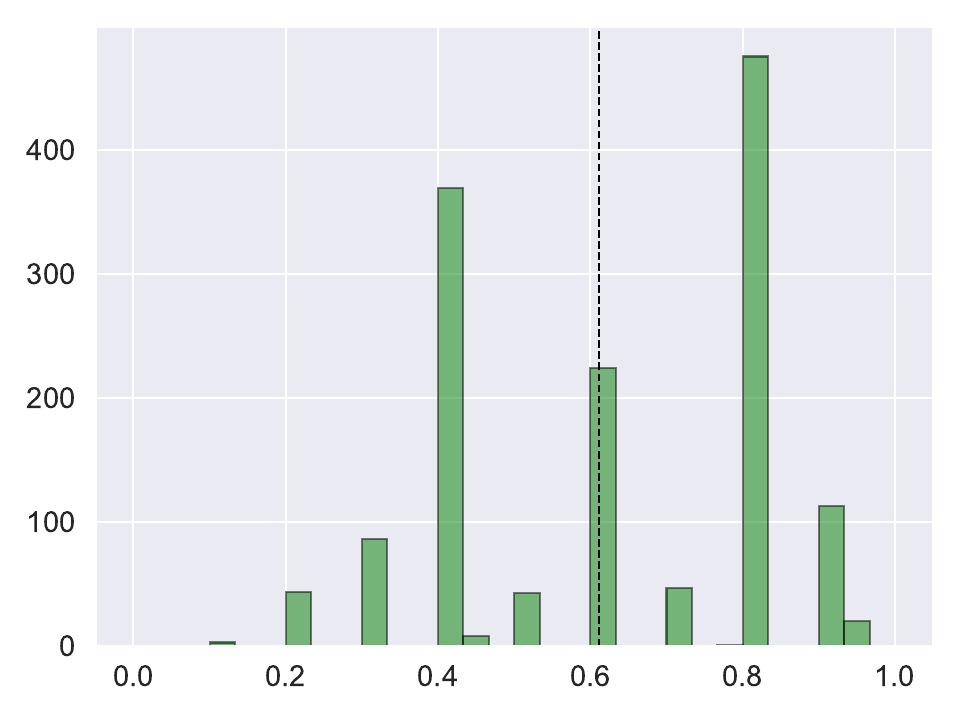}
    }
    \subfigure[Sentiment, LLaMA3, \textbf{ECE: 0.5714}]{
    \includegraphics[scale=0.26]{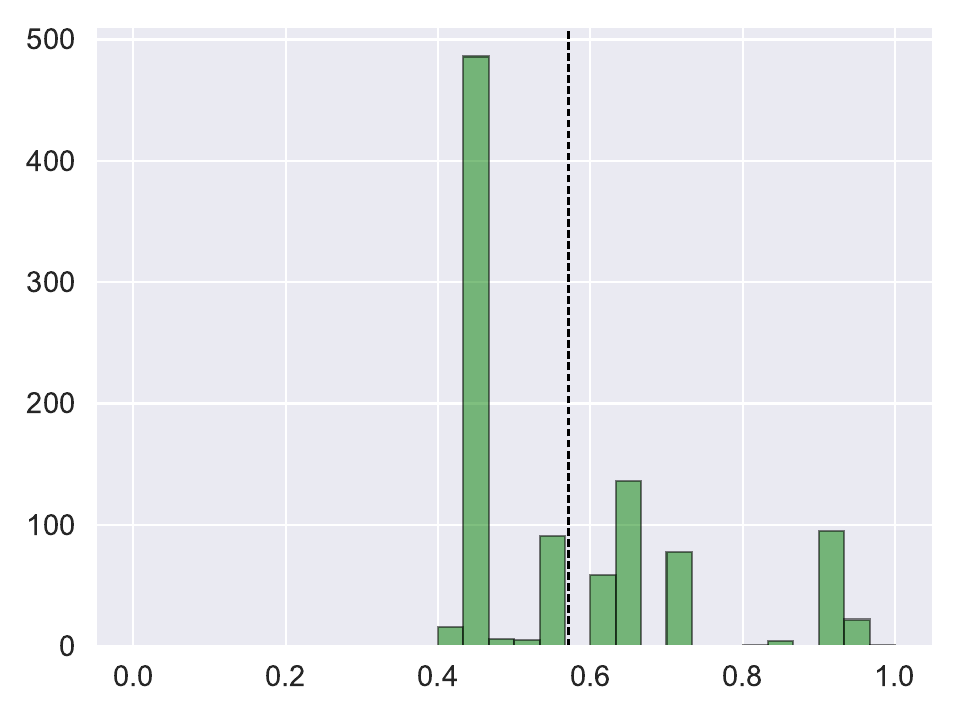}%
    }
    \subfigure[CD, GPT3.5, \textbf{ECE: 0.1915}]{
    \includegraphics[scale=0.26]{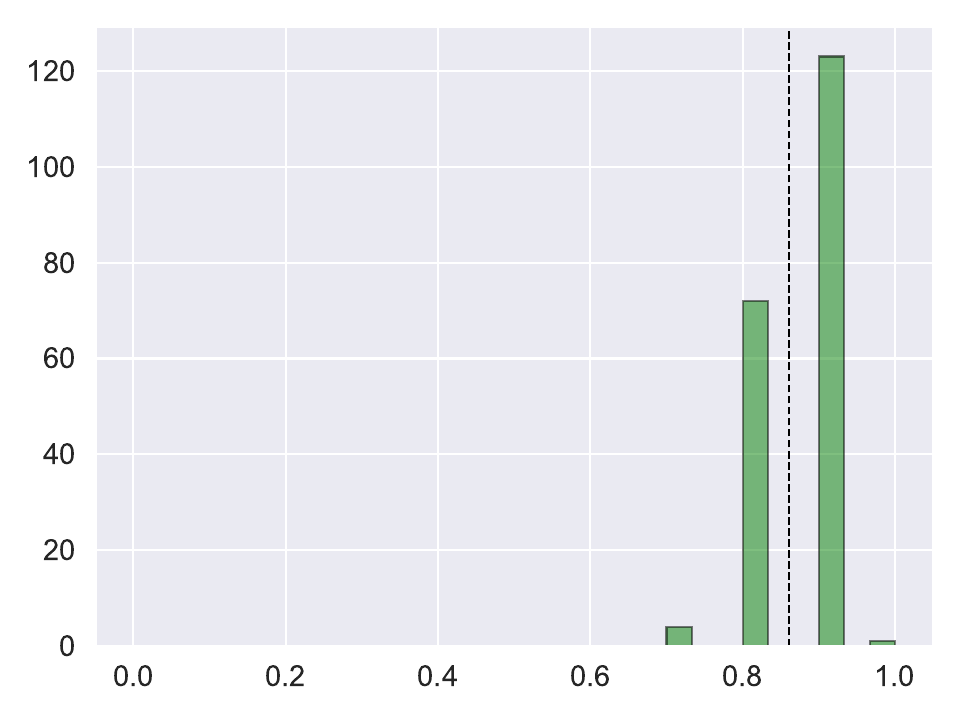}%
    }
    \subfigure[PC, GPT3.5, \textbf{ECE: 0.4560}]{
    \includegraphics[scale=0.26]{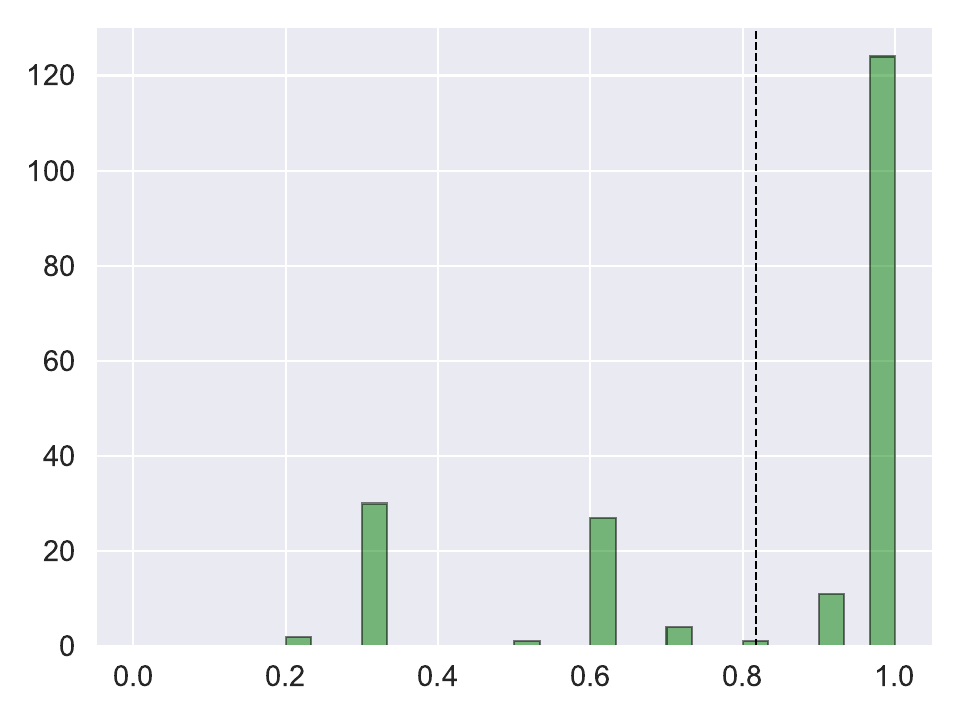}%
    }
    \subfigure[TMN, GPT3.5, \textbf{ECE: 0.0982}]{
    \includegraphics[scale=0.26]{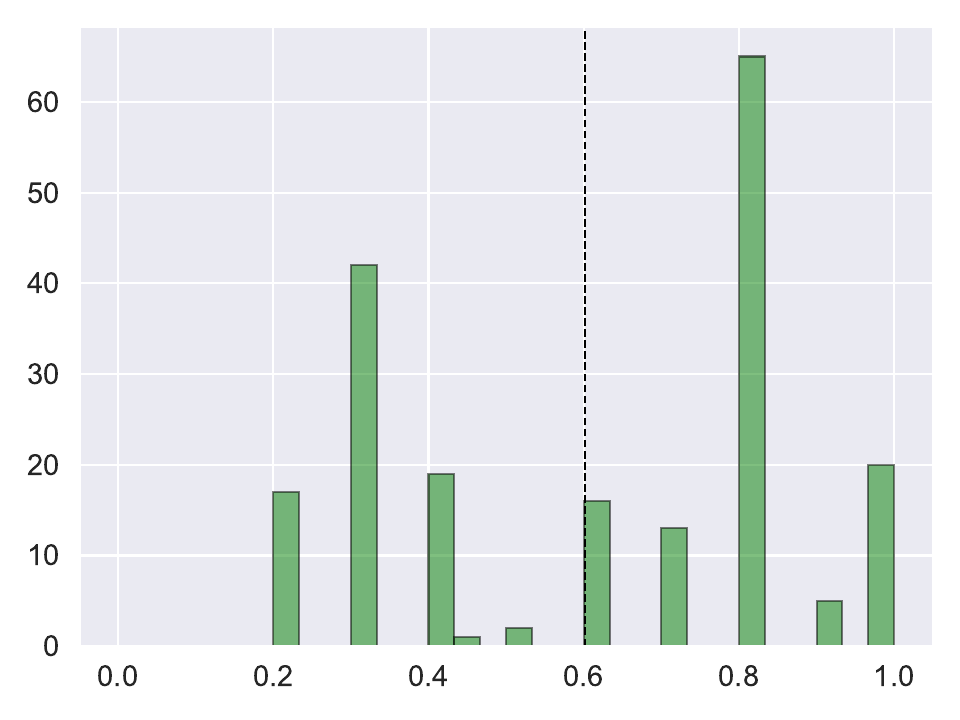}%
    }
    \subfigure[Sentiment, GPT3.5, \textbf{ECE: 0.0917}]{
    \includegraphics[scale=0.26]{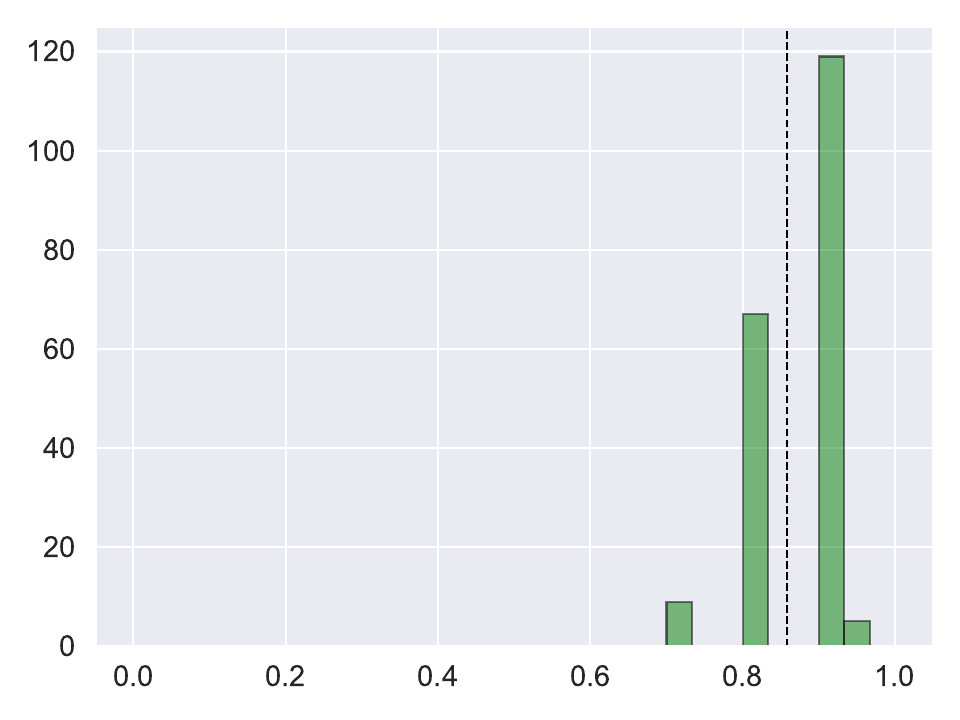}%
    }
    \subfigure[CD, GPT4, \textbf{ECE: 0.1630}]{
    \includegraphics[scale=0.26]{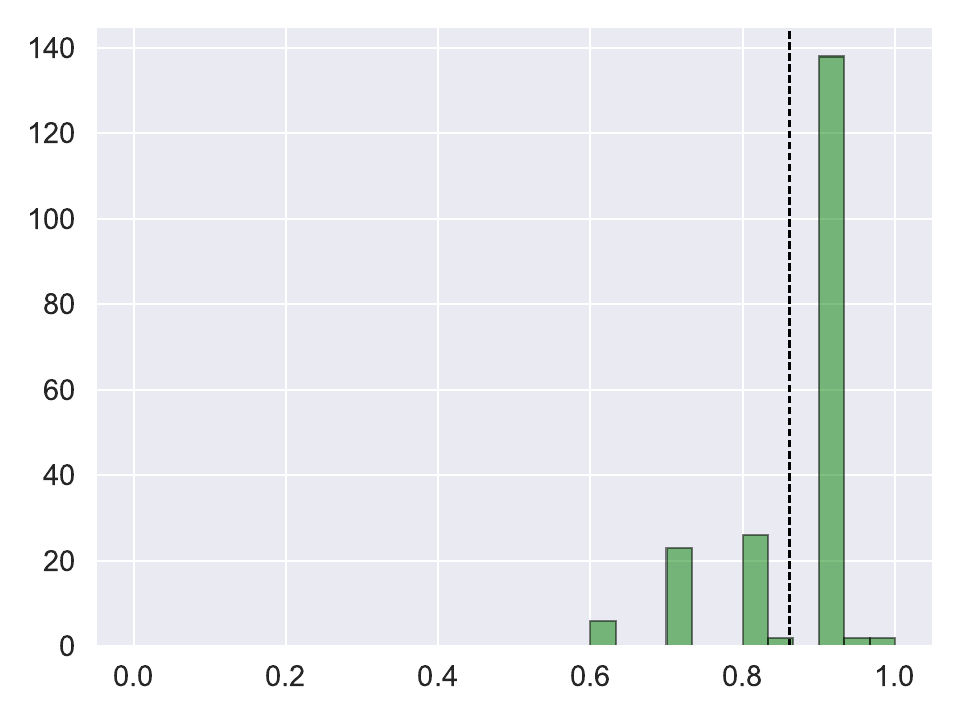}%
    }
    \subfigure[PC, GPT4, \textbf{ECE: 0.0143}]{
    \includegraphics[scale=0.26]{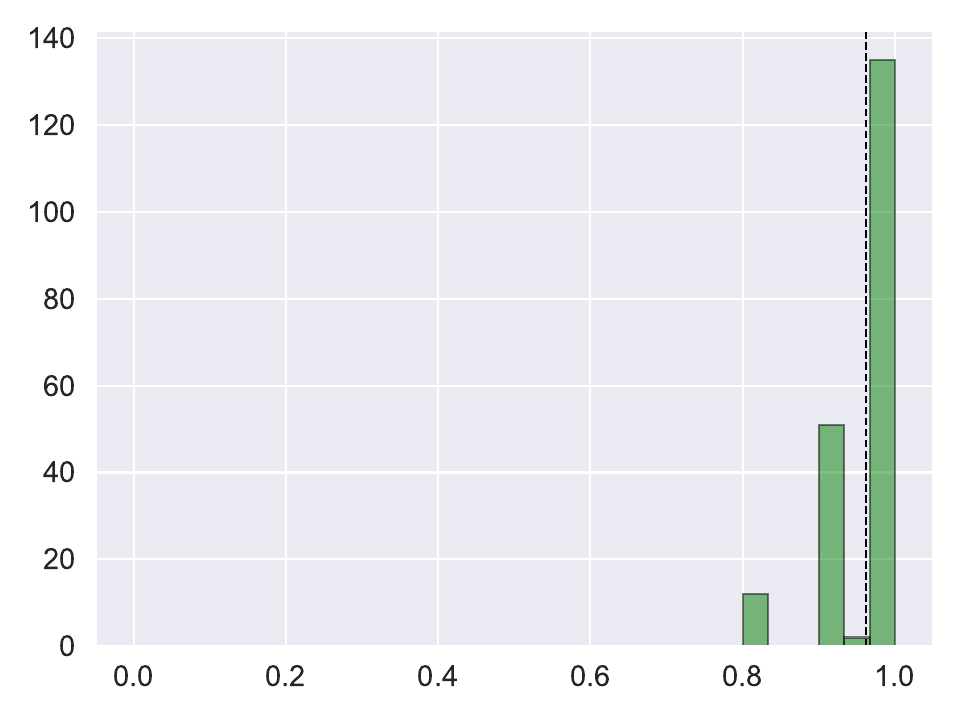}%
    }
    \subfigure[TMN, GPT4, \textbf{ECE: 0.0407}]{
    \includegraphics[scale=0.26]{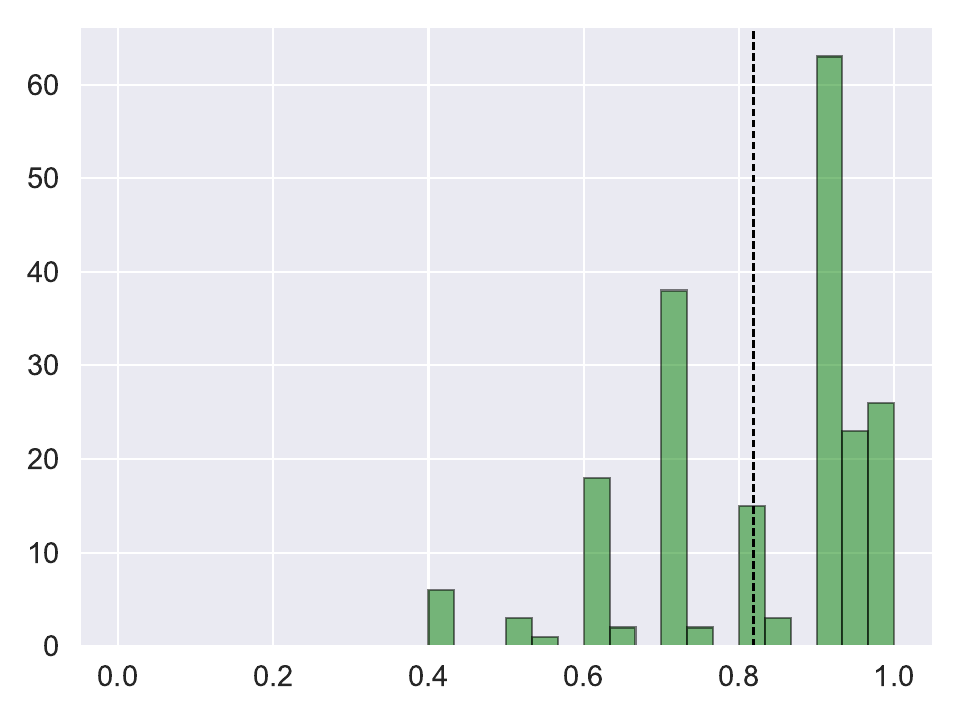}%
    }
    \subfigure[Sentiment, GPT4, \textbf{ECE: 0.0787}]{
    \includegraphics[scale=0.26]{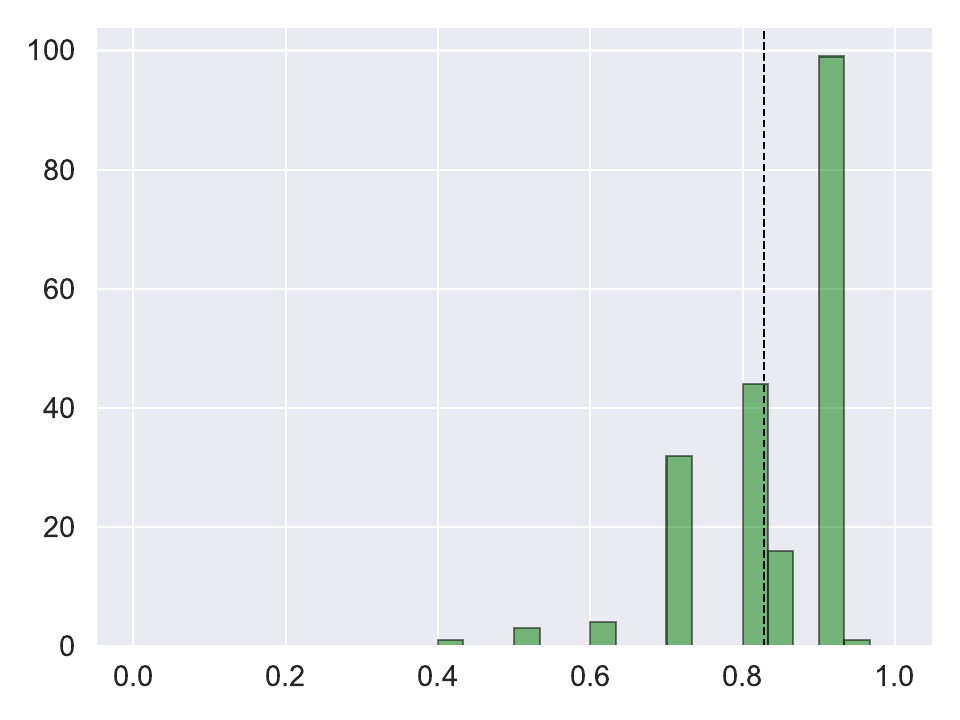}%
    }

    \caption{Distribution of prediction confidence and ECE for LLMs. The dashed line indicates the average confidence.}
    \label{fig:rq1_confidence}
\end{figure*}

First, as most fault detection methods rely on prediction confidence, we check whether LLMs can confidently predict the data in our studied tasks. Figure~\ref{fig:rq1_confidence} depicts the distribution~(number of data in each confidence interval) of the prediction confidence produced by LLMs, where we set the number of intervals $M$ as 30. The first conclusion is that GPT4 has the most confidence in its predictions compared to other LLMs. On average, the confidence of DeepSeekCoder, LLaMA3, GPT3.5, and GPT4 on these four datasets are 81.72\%, 59.15\%, 77.01\%, and 87.27\%, respectively. Comparing the accuracy of LLMs and their average confidence~(which is a notion of miscalibration~\cite{guo2017calibration}), we found that our LLMs are not well-calibrated to these four tasks. For example, for DeepSeekCoder on the clone detection task, the accuracy and average confidence are 69.4\% and 94.61\%, respectively. This means DeepSeekCoder is overconfident with its prediction even though it has poor performance. Overall, except for TagMyNews, LLMs are overconfident in the other three tasks with 16.78\% higher confidence than the true accuracy. Additionally, the results show that prediction confidence is concentrated in certain ranges for many cases. For example, for the clone detection - GPT3.5, confidence only falls into four ranges. This means the model has similar confidence to many different inputs, in turn, the output-based fault detection methods~(e.g., MaxP) could produce similar uncertainty scores to these inputs.

Interestingly, we found that GPT4 has perfect accuracy~(100\%) on the problem classification task with high average confidence~(0.9827 on average) and a low ECE score of 0.0143. We conjecture there is a data leakage problem of our test data, i.e., our used problem classification dataset is a part of the training data of GPT4. Investigating the exact reason behind this could be future work to better understand what has been learned by the pre-trained LLMs. As there are no faults in GPT4 for the problem classification task, we exclude this model from the following fault detection study.

\noindent
\\
{\framebox{\parbox{0.96\linewidth}{
\textbf{Answer to RQ1}: LLMs are not well-calibrated and overconfident in clone detection, problem classification, and news classification tasks. Besides, prediction confidence is concentrated in a few intervals, indicating LLMs have similar confidence to most of the inputs. }}}

\subsection{RQ2: Effectiveness of Fault Detection}
\label{sec:rq2}

In this RQ, we explore the performance of existing fault detection methods for LLMs. As mentioned in Section~\ref{sec:datasets}, we exclude models that have perfect performance~(100\% accuracy) on the studied tasks.

\begin{figure*}[h]
    \centering
    \subfigure[CD, DeepSeekCoder]{
    \includegraphics[scale=0.2]{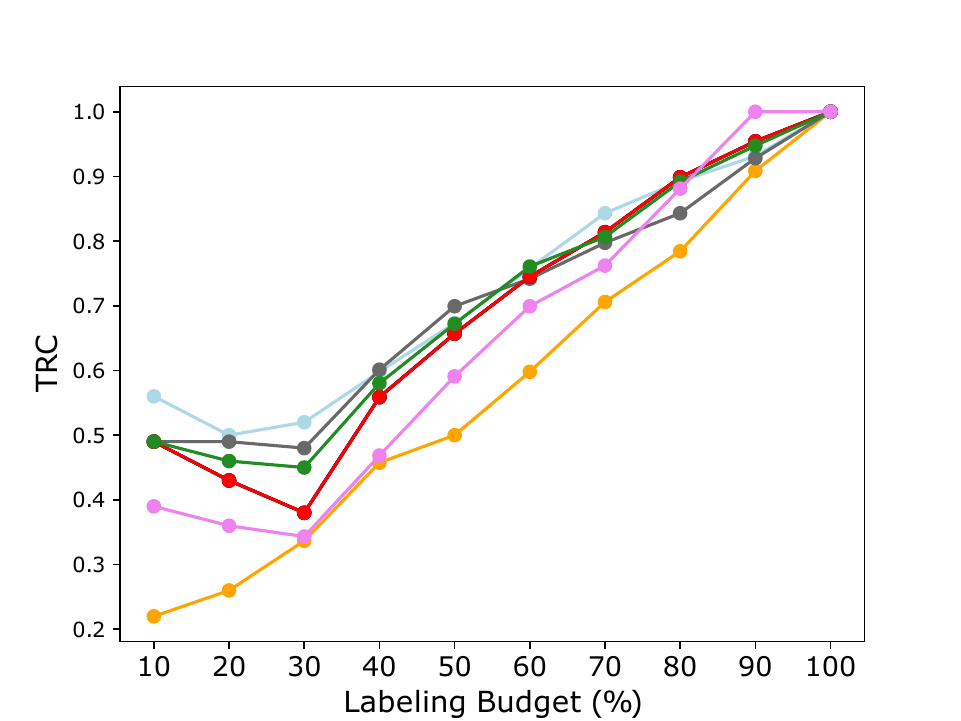}
    }
    \subfigure[PC, DeepSeekCoder]{
    \label{fig:pc_llama}
    \includegraphics[scale=0.2]{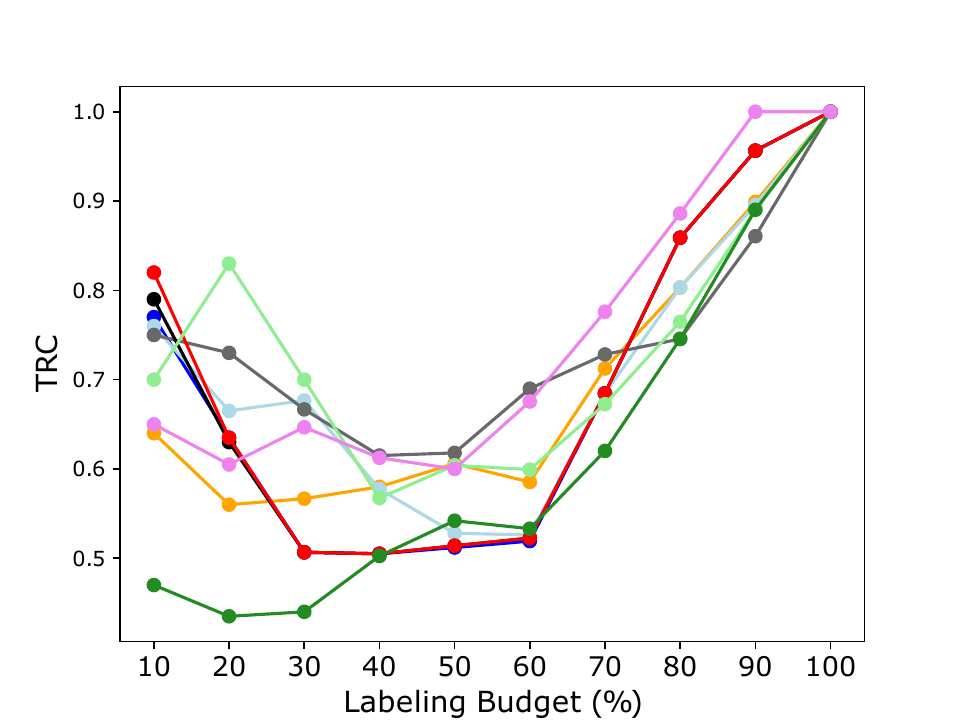}%
    }
    \subfigure[TMN, LLaMA3]{
    \label{fig:tmn_llama}
    \includegraphics[scale=0.2]{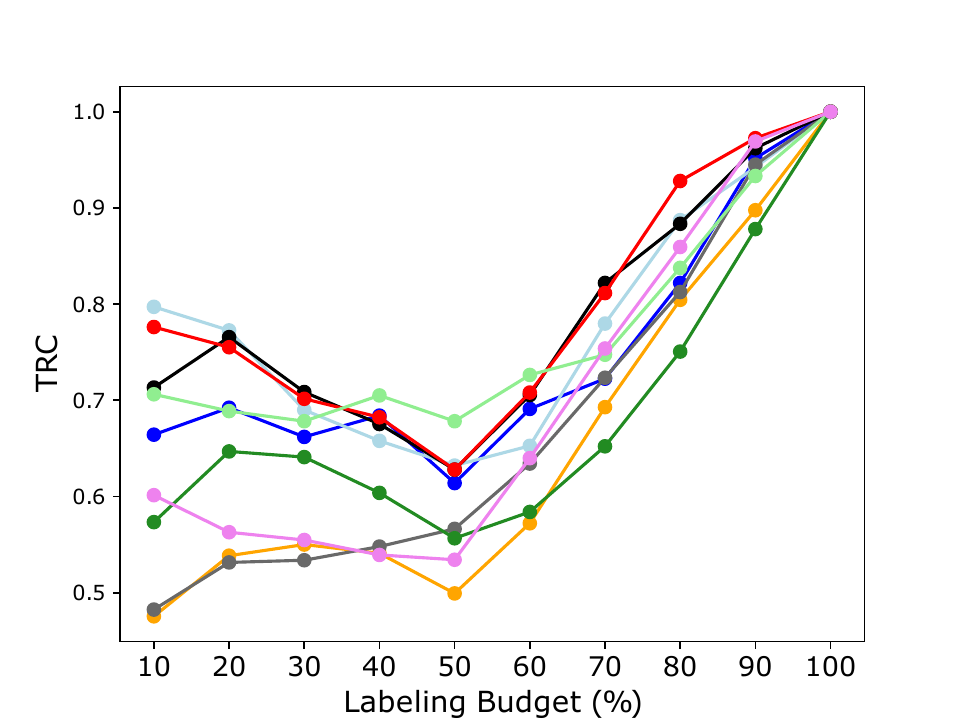}
    }
    \subfigure[Sentiment, LLaMA3]{
    \includegraphics[scale=0.2]{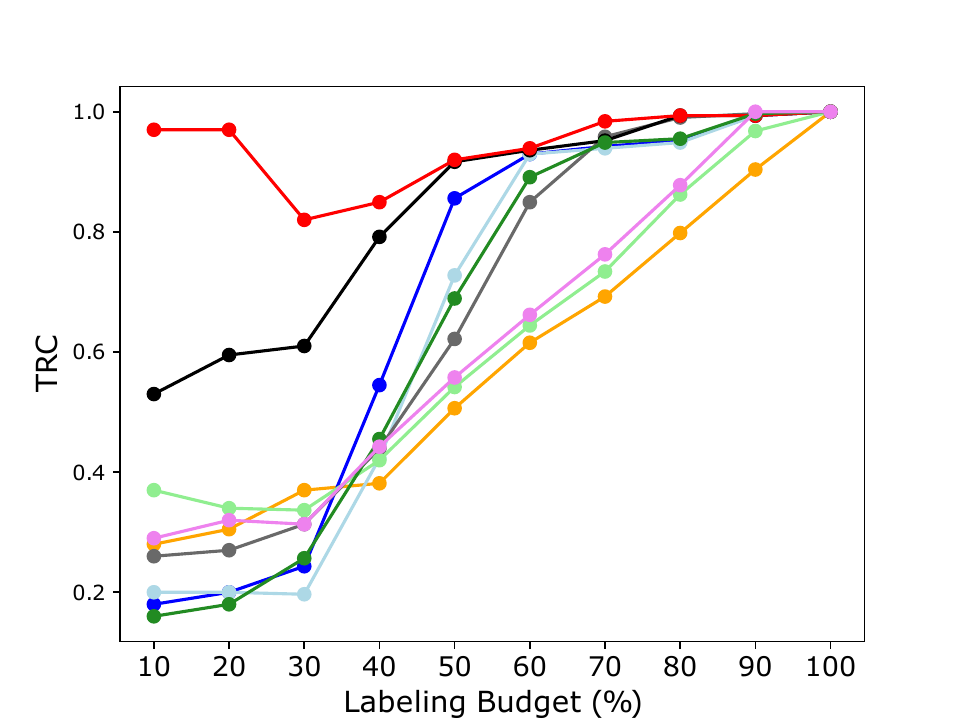}%
    }
    \subfigure[CD, GPT3.5]{
    \includegraphics[scale=0.2]{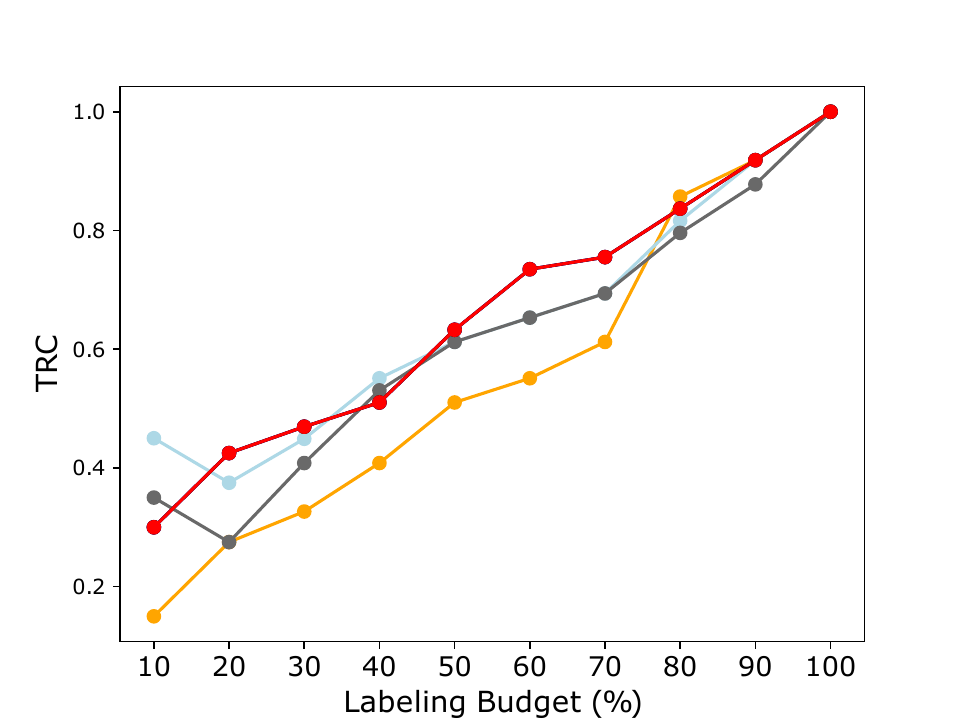}%
    }
    \subfigure[PC, GPT3.5]{
    \includegraphics[scale=0.2]{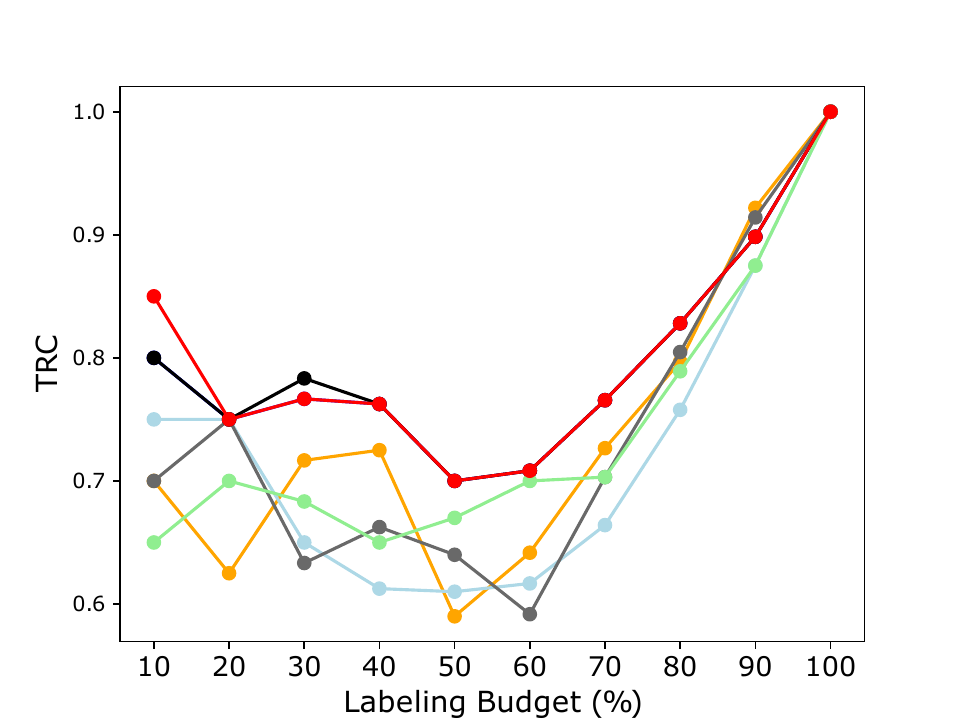}%
    }
    \subfigure[TMN, GPT3.5]{
    \includegraphics[scale=0.2]{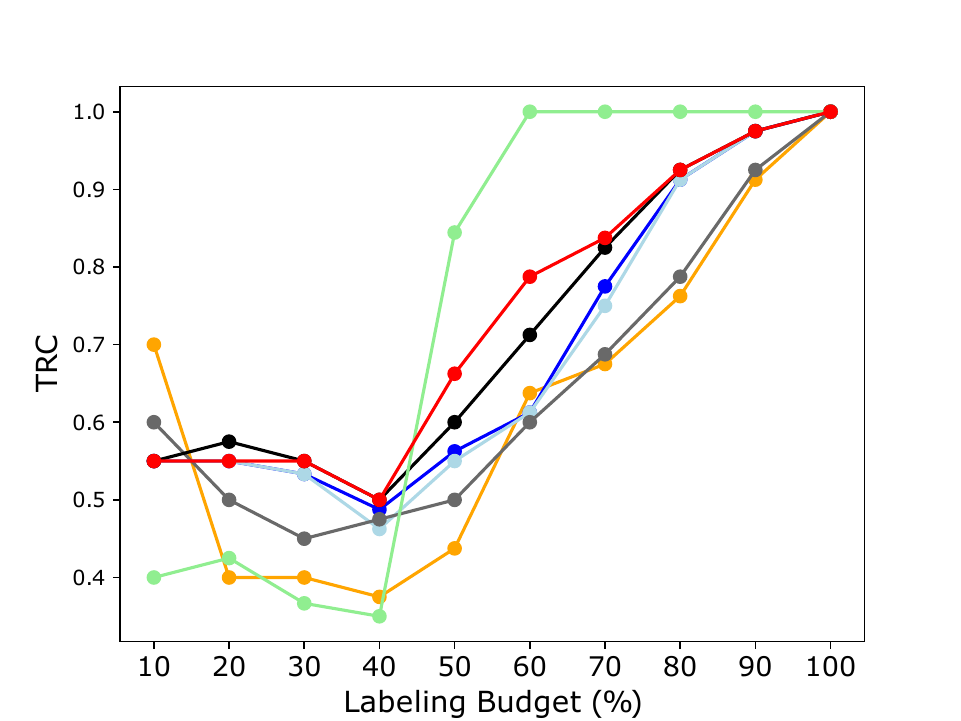}%
    }
    \subfigure[Sentiment, GPT3.5]{
    \includegraphics[scale=0.2]{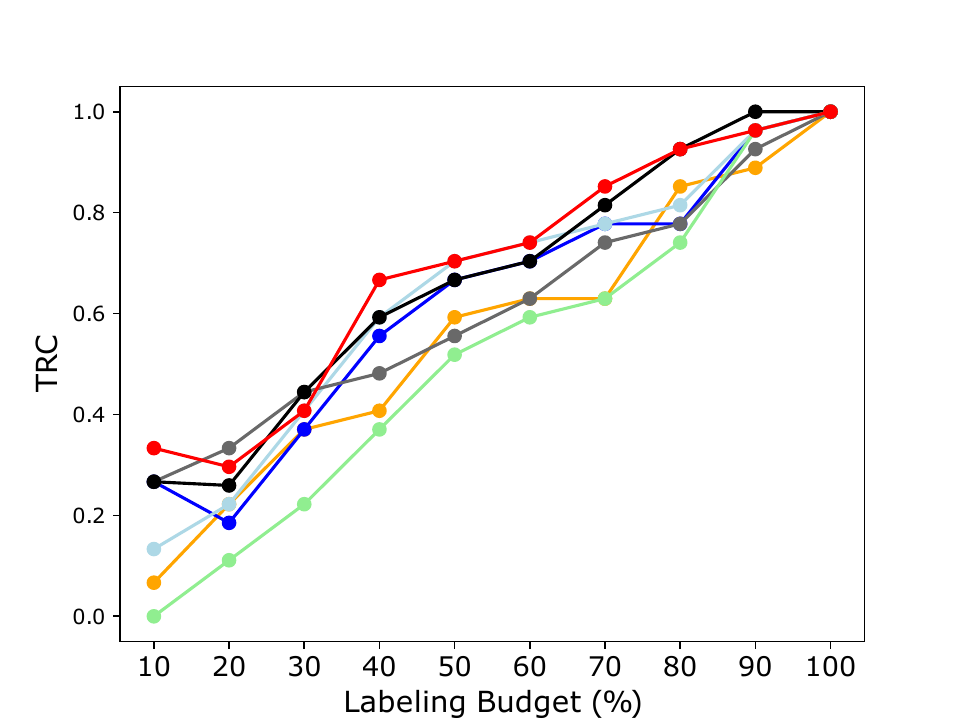}%
    }
    \subfigure[CD, GPT4]{
    \includegraphics[scale=0.2]{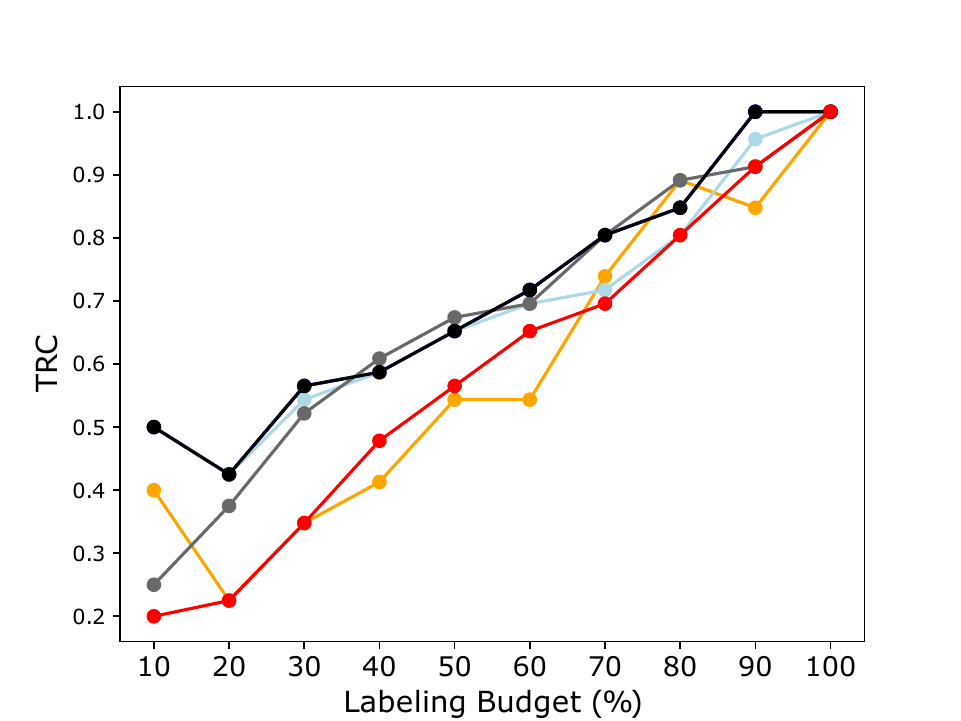}%
    }
    \subfigure[TMN, GPT4]{
    \includegraphics[scale=0.2]{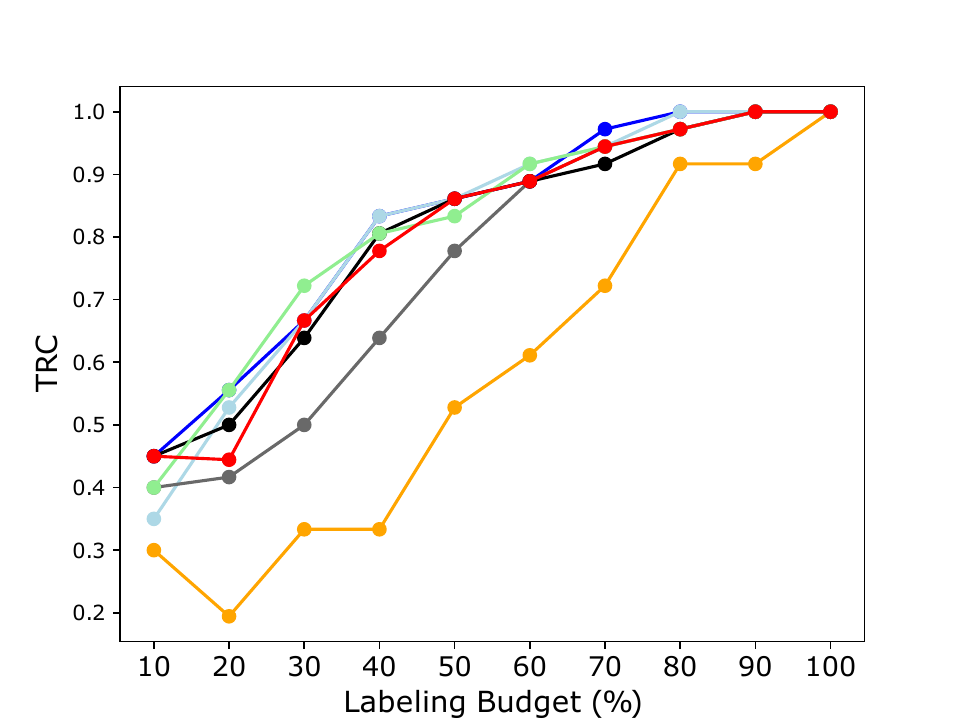}%
    }
    \subfigure[Sentiment, GPT4]{
    \includegraphics[scale=0.2]{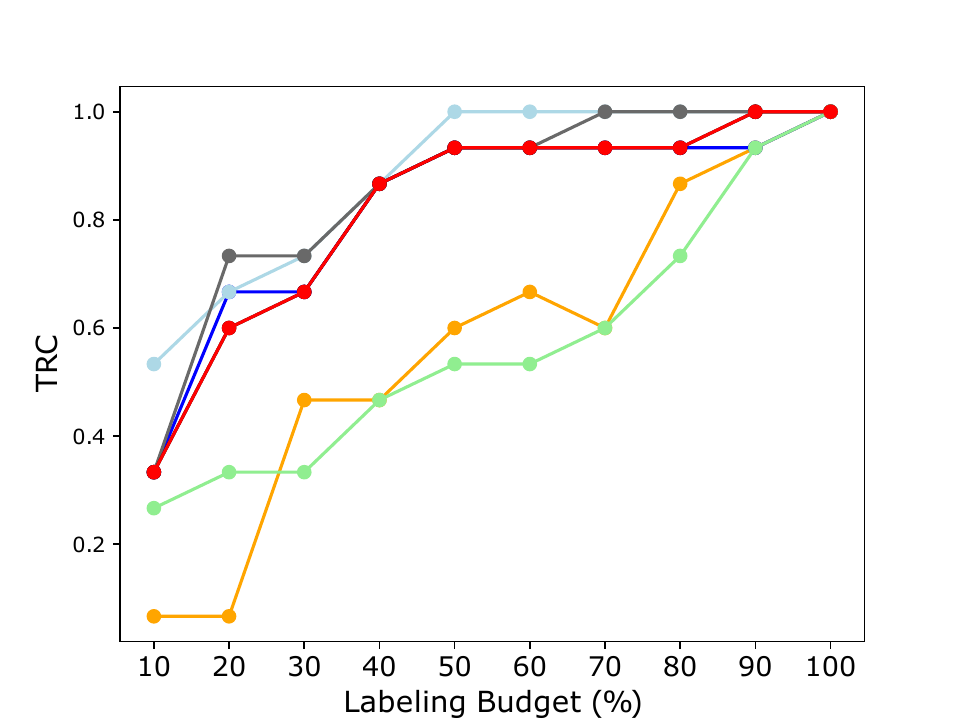}%
    }
    \subfigure[Legend]{
    \includegraphics[scale=0.41]{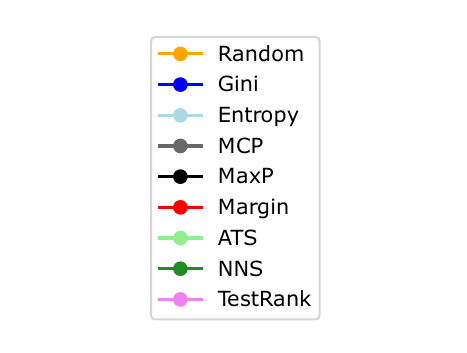}%
    }
    \caption{Test relative coverage. }
    \label{fig:rq2_trc_fig}
\end{figure*}

\begin{table*}[]
\caption{Average test relative coverage of each method. The best method is highlighted in \colorbox{mycolor_green}{green}.}
\label{tab:rq2_atrc}
\centering
\resizebox{.9\textwidth}{!}{
\begin{tabular}{lccccccccc}
 \bottomrule
 & \textbf{Random} & \multicolumn{1}{l}{\textbf{Gini}} & \multicolumn{1}{l}{\textbf{Entropy}} & \multicolumn{1}{l}{\textbf{Mcp}} & \multicolumn{1}{l}{\textbf{Maxp}} & \multicolumn{1}{l}{\textbf{Margin}} & \multicolumn{1}{l}{\textbf{ATS}} & \multicolumn{1}{l}{\textbf{NNS}} & \multicolumn{1}{l}{\textbf{TestRank}} \\ \bottomrule
DeepSeekCoder-CD & 0.5301 & 0.6586 & \cellcolor{mycolor_green}0.6973 & 0.6746 & 0.6586 & 0.6586 & - & 0.6732 & 0.6106 \\
GPT3.5-CD & 0.5121 & \cellcolor{mycolor_green}0.6202 & 0.6132 & 0.5774 & \cellcolor{mycolor_green}0.6202 & \cellcolor{mycolor_green}0.6202 & - & - & - \\
GPT4-CD & 0.5501 & \cellcolor{mycolor_green}0.6777 & 0.6535 & 0.6371 & 0.5424 & \cellcolor{mycolor_green}0.6777 & - & - & - \\
DeepSeekCoder-PC & 0.6614 & 0.6609 & 0.6796 & 0.7116 & 0.6631 & 0.6670 & 0.7031 & 0.5754 & \cellcolor{mycolor_green}0.7169 \\
GPT3.5-PC & 0.7160 & 0.7755 & 0.6984 & 0.7110 & 0.7774 & \cellcolor{mycolor_green}0.7811 & 0.7134 & - & - \\
LLaMA3-TMN & 0.6191 & 0.7227 & 0.7569 & 0.6419 & 0.7627 & \cellcolor{mycolor_green}0.7737 & 0.7446 & 0.6541 & 0.6683 \\
GPT3.5-TMN & 0.5889 & 0.6620 & 0.6551 & 0.6139 & 0.6903 & 0.7042 & \cellcolor{mycolor_green}0.7096 & - & - \\
GPT4-TMN & 0.5395 & \cellcolor{mycolor_green}0.8031 & 0.7889 & 0.7265 & 0.7815 & 0.7784 & 0.7944 & - & - \\
LLaMA3-Sentiment & 0.5392 & 0.6490 & 0.6176 & 0.6332 & 0.8131 &\cellcolor{mycolor_green} 0.9377 & 0.5796 & 0.6147 & 0.5806 \\
GPT3.5-Sentiment & 0.5177 & 0.5852 & 0.5951 & 0.5728 & 0.6305 & \cellcolor{mycolor_green}0.6543 & 0.4609 & - & - \\
GPT4-Sentiment & 0.5259 & 0.8000 & \cellcolor{mycolor_green}0.8667 & 0.8370 & 0.8000 & 0.8000 & 0.5259 & - & - \\ \bottomrule
Average & 0.5727 & 0.6923 & 0.6929 & 0.6670 & 0.7036 & \cellcolor{mycolor_green}0.7321 & 0.6540 & 0.6294 & 0.6441 \\ \bottomrule
\end{tabular}
}
\end{table*}

Figure~\ref{fig:rq2_trc_fig} depicts the TRC values achieved by fault detection methods in each labeling budget, and Table~\ref{tab:rq2_atrc} summarizes the averaged TRC values across all considered labeling budgets. First, considering the averaged results, we can see that all methods have relatively better results than random selection, which indicates the potential of their usefulness. However, we found that the best averaged TRC value on LLMs is only 0.7321, which is far away from the idea detection performance, TRC = 1. Worsely, in some cases, e.g., Sentiment-LLaMA3, some methods have less than 0.2 TRC score, which never happened in conventional deep learning models~\cite{NEURIPS2021_ae785101}. This means fault detection for LLMs is a more challenging problem compared to fault detection for conventional deep learning models.


Then, comparing each method, interestingly, we found that the simple method, Margin, which just utilizes the top-1 and top-2 output probabilities to measure the uncertainty, performs the best among our considered methods. This phenomenon is similar to the findings by existing works~\cite{weiss2022simple} which stated that simple methods perform the best for neural network test prioritization. Specifically, in more than half of cases~(6 out of 11 cases), Margin achieved the highest averaged TRC scores. Methods that require embedding information for the data selection~(NNS and TestRank) cannot stand out in LLMs. The reason could be that the embeddings extracted by LLMs are too high-dimensional and diverse to be learned by simple clustering methods~(used by NNS) and graph neural networks~(used by TestRank).


\noindent
\\
{\framebox{\parbox{0.96\linewidth}{
\textbf{Answer to RQ2}: Under our studied objects, the simple method, Margin, performs the best in fault detection for LLMs. However, there is still a big room for improvement, and more effective methods are needed.}}}

\section{MuCS: Mutation-Based Confidence Smoothing}
\label{sec:mutation}

\begin{wrapfigure}{R}{0.5\textwidth}
\begin{minipage}{0.5\textwidth}
\vspace{-5mm}
\begin{algorithm}[H] 
\small
\caption{MuCS}
\label{alg:confidence_smooth}
\SetAlgoLined
\Input{$p$: input prompt\\
$OPs$: mutation operators\\
$M$: LLM\\
$K$: perturbation size\\
$n$: number of generated mutants\\
}
\Output{$C_{smooth}$: smoothed confidence}
$Cs = []$ \\
\For{$i=0 \to n$}
{
\tcc{{\footnotesize{mutants generation}}}
    \For{$j=0 \to K$} 
    {
    $OP = \mathsf{RandomSelection}\left(OPs\right)$ \\
    $p' = OP\left(p\right)$ \\
    $p = p'$ \\
    }
\tcc{{\footnotesize{confidence calculation using Definition~\ref{def:confidence}}}}
$Cs.append\left(\mathsf{Confidence}\left(M\left(p\right)\right)\right)$ 
}
\tcc{{\footnotesize{confidence averaging}}}
$C_{smooth} = \mathsf{Mean}\left(Cs\right)$ \\
\Return $Cs$;
\end{algorithm}
\end{minipage}
\end{wrapfigure}

As analyzed in Section~\ref{sec:rq1}, the concentrated confidence issue can harm the performance of output-based fault detection methods~(those methods will give similar uncertainty scores to the inputs), we need to diversify the prediction confidence of LLMs to enhance fault detection. It is challenging since we cannot get the internal information~(e.g., output logits) of closed-source LLMs to smooth the prediction confidence.  Inspired by previous works~\cite{10.1109/ICSE43902.2021.00046, 10.1109/ICSE48619.2023.00110} that showed the potential of testing deep learning models by mutation analysis~(one uses mutation killing score to detect faults and the other one uses mutation analysis to improve the performance of code models), we propose MuCS, a simple yet effective black-box framework to enhance existing fault detection methods by smoothing the prediction confidence from the input-level using prompt mutation. 

\begin{figure*}[]
    \centering
    \includegraphics[scale=0.4]{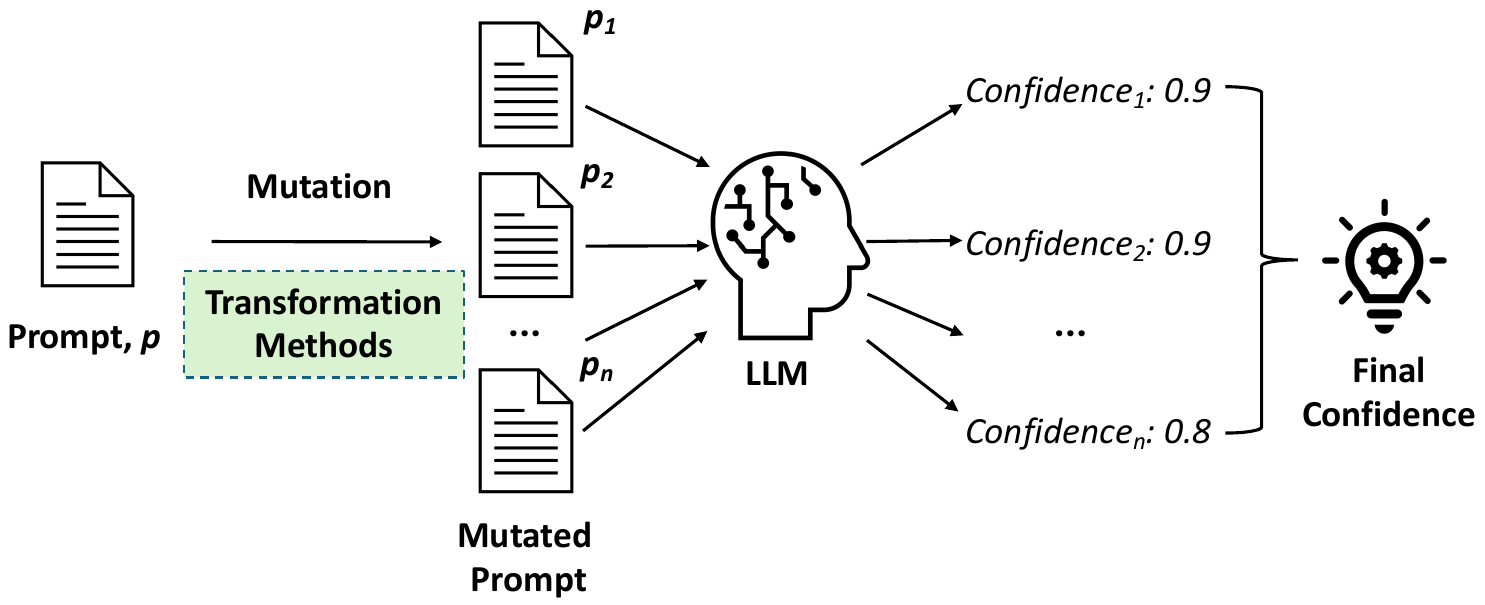}
    \caption{Workflow of MuCS.}
    \label{fig:mutation}
\end{figure*}

Figure~\ref{fig:mutation} presents the overall workflow of our proposed framework. It has two main steps, 1) prompt mutation, and 2) confidence smooth. Specifically, given an LLM and an input prompt \textit{p}, we first utilize transformation methods to mutate the inputs and generate \textit{n} mutated prompts, $p_{1}$, $p_{2}$, ..., $p_{n}$. For NLP tasks, we use text augmentation methods~\cite{marivate2020improving}~(e.g., token random deletion) to transform prompts into new ones. For the code tasks, we consider both text augmentation methods and code refactoring methods~\cite{al2017empirical}~(e.g., add new \textit{Print()} functions) for the prompt mutation. After that, we feed all the mutants to the LLM and collect the output confidence $Confidence_{1}$, $Confidence_{2}$,\ldots, $Confidence_{n}$ produced by the LLM. Finally, a confidence smooth method is employed to compute the \textit{final confidence} of the LLM on the input. In this work, we use the straightforward method, average, to smooth the confidence. Algorithm~\ref{alg:confidence_smooth} summarizes our mutation-based confidence smoothing solution. MuCS first randomly selects multiple (defined by the perturbation size $K$) mutation operators $OP$~(line 4) to mutate $p$ and then generates $p'$~(lines 4, 5). Then, MuCS collects all the prediction confidence produced by LLMs using all mutants $p'$~(line 8). Finally, the averaged confidence is returned as the smoothed confidence~(lines 10 and 11).

\subsection{Mutation Operators}
Two types of mutation operators have been included in the framework, text augmentation methods and code refactoring methods. In total, five text augmentation methods~(including Synonym Replacement, Random Deletion, Random Insertion, Random Swap, and Punctuation Insertion), 
and four code refactoring methods~(including Print Adding, Local Variable Adding, Dead If Adding, and Duplication) have been used in MuCS. Given a prompt $p$ with the length of works $k$, each mutation operator is defined as follows.

\begin{itemize}[leftmargin=*]
    \item \textbf{Synonym Replacement~(SP)} randomly selects $n$ words in $p$ and replaces them as their synonyms, where $n < k$. Here, the synonyms are searched from the WordNet~\cite{10.1145/219717.219748}. We use the default setting of TextAugment~\cite{marivate2020improving}, $n = 1$ in MuCS.

    \item  \textbf{Random Deletion~(RD)} randomly removes words in the $p$ with a probability. Specifically, RD assigns a random probability $pro$ uniformed from 0 to 1 to each word in $p$. Given a probability threshold $T$, if $pro$ is less than $T$, the corresponding word will be removed. We use the default setting of MixCode~\cite{dong2023mixcode}, $T = 0.01$ in MuCS.

    \item \textbf{Random Insertion~(RI)} randomly selects $n$ words in $p$ first where $n < k$. Then, it finds the synonyms of selected words and inserts them into the random locations of $p$. We use the default setting of MixCode, $n = 1$ in MuCS.

    \item \textbf{Random Swap~(RW)} randomly swaps two words in $p$. 

    \item  \textbf{Punctuation Insertion~(PI)} randomly selects $n$ punctuation~(e.g., !) and inserts them into the random locations of $p$, where $n < k$. We use the default setting of MixCode, $n = 1$ in MuCS.

    \item  \textbf{Print Adding~(PA)} randomly selects a line in the code and adds a print function with randomly generated content below it. 
    \item  \textbf{Local Variable Adding~(LVA)} randomly selects a line in the code and defines an unused local variable below it.
    \item  \textbf{Dead If Adding~(DIA)} randomly selects a line in the code and then adds an unreachable if statement below it.
    \item  \textbf{Duplication} randomly copies an assignment in the code and inserts it next to this assignment.  
\end{itemize}

During the mutant generation process, we randomly select $K$ mutation operators to mutate $p$ and generate its mutant $p'$.

\section{Evaluation}
\label{sec:mutation_evaluation}

\subsection{Research Question}

To investigate whether our proposed framework MuCS can help fault detection for LLMs, we first analyze the effectiveness of existing fault detection with MuCS as a plug-in. Then, we explore the reason why MuCS enhances fault detection by analyzing its influences on the prediction confidence of LLMs. Our exploration plans to answer the following research questions.

\begin{itemize}[leftmargin=*]

    \item  \textbf{RQ3: How effective is MuCS in enhancing fault detection for LLMs?}  In this research question, we compare the effectiveness of fault detection methods with and without using MuCS to analyze the usefulness of our proposed framework. 
    
    \item \textbf{RQ4: How does MuCS affect the final outputs of the LLMs?} We want to explore how MuCS influences the outputs of LLMs to figure out why it works on fault detection.

\end{itemize}

\subsection{Evaluation Setup}

We set the mutation times as 10 for all tasks. That is, we generate 10 mutants for each input prompt.  We set $K$~(the number of mutation operators used) as 3 for all the tasks except for the clone detection. The reason is that we found the performance of LLMs on the clone detection task drops significantly~(nearly random guesses, e.g., all the predicted labels of GPT3.5 are 0) when $K = 3$ and $K = 2$. Therefore, we set $K = 1$ for the clone detection task. For other tasks, LLMs have similar accuracy on the mutated datasets to the original ones~(with less than 5\% difference). For the other settings, we use the same settings as Section~\ref{sec:design}.

\subsection{RQ3: Effectiveness of MuCS}

\begin{table*}[]
\caption{Averaged test relative coverage (relative improvement compared to results without using MuSC in Table~\ref{tab:rq2_atrc}) cross all labeling budgets of each method after prompt mutation-based confidence smoothing. }
\label{tab:rq3atrc}
\centering
\resizebox{1.\textwidth}{!}{
\begin{tabular}{llllllllll}
\bottomrule
 & \textbf{Gini-M} & \textbf{Entropy-M} & \textbf{MCP-M} & \textbf{MaxP-M} & \textbf{Margin-M} & \textbf{ATS-M} & \textbf{NNS-M} & \textbf{TestRank-M} & \textbf{BALD} \\ \bottomrule
DeepSeekCoder-CD & 0.6886 ($\textcolor{green}{\uparrow}$4.55) & 0.6890 ($\textcolor{red}{\downarrow}$1.20) & 0.6592 ($\textcolor{red}{\downarrow}$2.28) & 0.6886 ($\textcolor{green}{\uparrow}$4.55) & 0.6886 ($\textcolor{green}{\uparrow}$4.55) & - & 0.6576 ($\textcolor{red}{\downarrow}$2.32) & 0.6134 ($\textcolor{green}{\uparrow}$0.46) & 0.7067 \\
GPT3.5-CD & 0.6873 ($\textcolor{green}{\uparrow}$10.81) & 0.5989 ($\textcolor{red}{\downarrow}$2.33) & 0.6369 ($\textcolor{green}{\uparrow}$10.31) & 0.6873 ($\textcolor{green}{\uparrow}$10.81) & 0.6873 ($\textcolor{green}{\uparrow}$10.81) & - & - & - & 0.5827 \\
GPT4-CD & 0.7550 ($\textcolor{green}{\uparrow}$11.41) & 0.7211 ($\textcolor{green}{\uparrow}$10.34) & 0.7318 ($\textcolor{green}{\uparrow}$14.87) & 0.7550 ($\textcolor{green}{\uparrow}$39.20) & 0.5107 ($\textcolor{red}{\downarrow}$24.64) & - & - & - & 0.4380 \\
DeepSeekCoder-PC & 0.7462 ($\textcolor{green}{\uparrow}$12.91) & 0.7134 ($\textcolor{green}{\uparrow}$4.97) & 0.6883 ($\textcolor{red}{\downarrow}$3.27) & 0.7412 ($\textcolor{green}{\uparrow}$11.77) & 0.6516 ($\textcolor{red}{\downarrow}$2.31) & 0.7212 ($\textcolor{green}{\uparrow}$2.57) & 0.6984 ($\textcolor{green}{\uparrow}$21.37) & 0.7425 ($\textcolor{green}{\uparrow}$3.57) & 0.7145 \\
GPT3.5-PC & 0.8344 ($\textcolor{green}{\uparrow}$7.59) & 0.8346 ($\textcolor{green}{\uparrow}$19.49) & 0.7116 ($\textcolor{green}{\uparrow}$0.08) & 0.8492 ($\textcolor{green}{\uparrow}$9.24) & 0.8343 ($\textcolor{green}{\uparrow}$6.81) & 0.7930 ($\textcolor{green}{\uparrow}$11.16) & - & - & 0.8389 \\
LLaMA3-TMN & 0.7637 ($\textcolor{green}{\uparrow}$5.68) & 0.7722 ($\textcolor{green}{\uparrow}$2.02) & 0.7073 ($\textcolor{green}{\uparrow}$10.18) & 0.7702 ($\textcolor{green}{\uparrow}$0.99) & 0.7653 ($\textcolor{red}{\downarrow}$1.09) & 0.7162 ($\textcolor{red}{\downarrow}$3.82) & 0.7156 ($\textcolor{green}{\uparrow}$9.41) & 0.7086 ($\textcolor{green}{\uparrow}$6.03) & 0.7434 \\
GPT3.5-TMN & 0.7431 ($\textcolor{green}{\uparrow}$12.24) & 0.7319 ($\textcolor{green}{\uparrow}$11.72) & 0.6208 ($\textcolor{green}{\uparrow}$1.13) & 0.7630 ($\textcolor{green}{\uparrow}$10.54) & 0.7481 ($\textcolor{green}{\uparrow}$6.24) & 0.7708 ($\textcolor{green}{\uparrow}$8.63) & - & - & 0.7407 \\
GPT4-TMN & 0.7691 ($\textcolor{red}{\downarrow}$4.23) & 0.7327 ($\textcolor{red}{\downarrow}$7.12) & 0.7451 ($\textcolor{green}{\uparrow}$2.55) & 0.7790 ($\textcolor{red}{\downarrow}$0.32) & 0.7821 ($\textcolor{green}{\uparrow}$0.48) & 0.7852 ($\textcolor{red}{\downarrow}$1.16) & - & - & 0.7037 \\
LLaMA3-Sentiment & 0.6417 ($\textcolor{red}{\downarrow}$1.13) & 0.5899 ($\textcolor{red}{\downarrow}$4.48) & 0.8423 ($\textcolor{green}{\uparrow}$33.02) & 0.7587 ($\textcolor{red}{\downarrow}$6.70) & 0.8643 ($\textcolor{red}{\downarrow}$7.83) & 0.6211 ($\textcolor{green}{\uparrow}$7.16) & 0.6006 ($\textcolor{red}{\downarrow}$2.30) & 0.6065 ($\textcolor{green}{\uparrow}$4.46) & 0.6199 \\
GPT3.5-Sentiment & 0.7654 ($\textcolor{green}{\uparrow}$30.80) & 0.7646 ($\textcolor{green}{\uparrow}$28.49) & 0.7959 ($\textcolor{green}{\uparrow}$38.94) & 0.7539 ($\textcolor{green}{\uparrow}$19.58) & 0.6782 ($\textcolor{green}{\uparrow}$3.65) & 0.7860 ($\textcolor{green}{\uparrow}$70.53) & - & - & 0.6379 \\
GPT4-Sentiment & 0.8593 ($\textcolor{green}{\uparrow}$7.41) & 0.8593 ($\textcolor{red}{\downarrow}$0.85) & 0.8444 ($\textcolor{green}{\uparrow}$0.88) & 0.8593 ($\textcolor{green}{\uparrow}$7.41) & 0.8519 ($\textcolor{green}{\uparrow}$6.49) & 0.7481 ($\textcolor{green}{\uparrow}$42.24) & - & - & 0.6667 \\ \bottomrule
Average & 0.7503 ($\textcolor{green}{\uparrow}$8.39) & 0.7280 ($\textcolor{green}{\uparrow}$5.05) & 0.7258 ($\textcolor{green}{\uparrow}$8.81) & 0.7641 ($\textcolor{green}{\uparrow}$8.60) & 0.7329 ($\textcolor{green}{\uparrow}$0.12) & 0.7427 ($\textcolor{green}{\uparrow}$13.57) & 0.6681 ($\textcolor{green}{\uparrow}$6.15) & 0.6678 ($\textcolor{green}{\uparrow}$3.67) & 0.6720 \\ \bottomrule
\end{tabular}
}
\end{table*}

\begin{table*}[]
\caption{Statistical analysis for comparing fault detection methods with and without using MuCS.}
\label{tab:rq3statistical}
\centering
\resizebox{.9\textwidth}{!}{
\begin{tabular}{llcccccccc}
\bottomrule
 & \multicolumn{1}{c}{} & \textbf{Gini} & \textbf{Entropy} & \textbf{Mcp} & \textbf{Maxp} & \textbf{Margin} & \textbf{ATS} & \textbf{NNS} & \textbf{TestRank} \\ \hline
\textbf{Welch’s t-test} & Significance & 2.8873 & 1.7373 & 3.0355 & 2.5787 & 1.1567 & 3.1049 & 1.1405 & 0.7018 \\
 & P-value & 4.10E-03 & 8.31E-02 & 2.56E-03 & 1.03E-02 & 2.48E-01 & 2.11E-03 & 2.56E-01 & 4.84E-01 \\ \hline
\textbf{Wilcoxon signed-rank test} & Significance & 500.5 & 1136.5 & 770.5 & 663.5 & 1273.5 & 283 & 148.5 & 42.5 \\
 & P-value & 4.82E-10 & 2.47E-04 & 1.30E-06 & 9.15E-09 & 1.18E-03 & 1.95E-06 & 1.09E-02 & 2.11E-05 \\ \hline
\end{tabular}
}
\end{table*}

First, we explore whether the MuCS can enhance the existing fault detection methods or not. Before that, since the input mutation can produce multiple predictions from LLMs which is similar to the dropout uncertainty~\cite{gal2016dropout}. We introduce another fault detection method baseline which is based on dropout uncertainty, Bayesian Active Learning by Disagreement~(BALD)~\cite{houlsby2011bayesian}.

\textbf{Bayesian Active Learning by Disagreement~(BALD)} defines the uncertainty of an input by the disagreement of the outputs produced by LLMs on the mutants of this input. The disagreement score is calculated by $1 - \frac{count\left(mode\left(y^1,...,y^T\right)\right)}{T} $, where $y_{i}$ is the predicted label of mutant $p_{i}$ and $T$ is the number of mutants. A higher BALD score indicates the input is more likely to be a fault.

Table~\ref{tab:rq3atrc} summarizes the average TRC scores achieved by different fault detection methods across all considered labeling budgets using MuCS~(with -M as a suffix), and the relative difference between the scores with and without using MuCS. The results clearly demonstrate that the MuCS significantly boosts the performance of fault detection. Specifically, in 52 out of 71 cases, the performance of existing fault detection methods has been increased with TRC improvements ranging from 0.48\% to 70.58\%. Considering the average results~(the last row in the Table), MuCS can enhance all fault detection methods with improvement by up to 13.57\%.  Besides, we conducted the statistical analysis using two methods Welch's t-test and Wilcoxon signed-rank test to investigate the significance of improvements introduced by MuCS. Concretely, we collect all TRC scores resulting from each method with and without using MuCS and compute the significance. The results are shown in Table~\ref{tab:rq3statistical}. The t-test suggests that for four methods~(Gini, MCP, MaxP, and ATS), the improvements are significant, and the signed-rank test demonstrates all improvements are significant~(with P-value < 0.03).   Therefore, we can conclude that MuCS is effective in boosting the existing fault detection methods on LLMs.

Comparing the improvements across different LLMs, we found that improvements in closed-source LLMs are greater than improvements in open-source LLMs. On average, the improvements in DeepSeekCoder, LLaMA3, GPT3.5, and GPT4 are 3.82\%, 3.22\%, 16.38\%, and 7.87\%, respectively. This means MuCS is an effective framework to enhance the black-box testing of LLMs. Then, after comparing each method after using MuCS, we found that MaxP, another simple method that directly utilizes the top-output probability as the indicator for data prioritization performs the best. Combining the conclusion in Section~\ref{sec:rq2}, simple methods perform better than learning-based methods with and without using MuCS.

Finally, even though the mutation-based confidence smooth can enhance the fault detection for LLMs, the best method only has a 0.7641 average TRC score. There is still a big room between the TRC scores achieved by existing methods and the ideal performance. 

\noindent
\\
{\framebox{\parbox{0.96\linewidth}{
\textbf{Answer to RQ3}: MuCS significantly enhances the effectiveness of existing fault detection methods by up to 70.53\% in terms of the TRC score. }}}

\subsection{RQ4: MuCS Affected Prediction Confidence}
\label{sec:answer2rq4}

\begin{figure*}[h]
    \centering

    \subfigure[CD, DeepSeekCoder, \textbf{ECE: 0.9112}]{
    \includegraphics[scale=0.26]{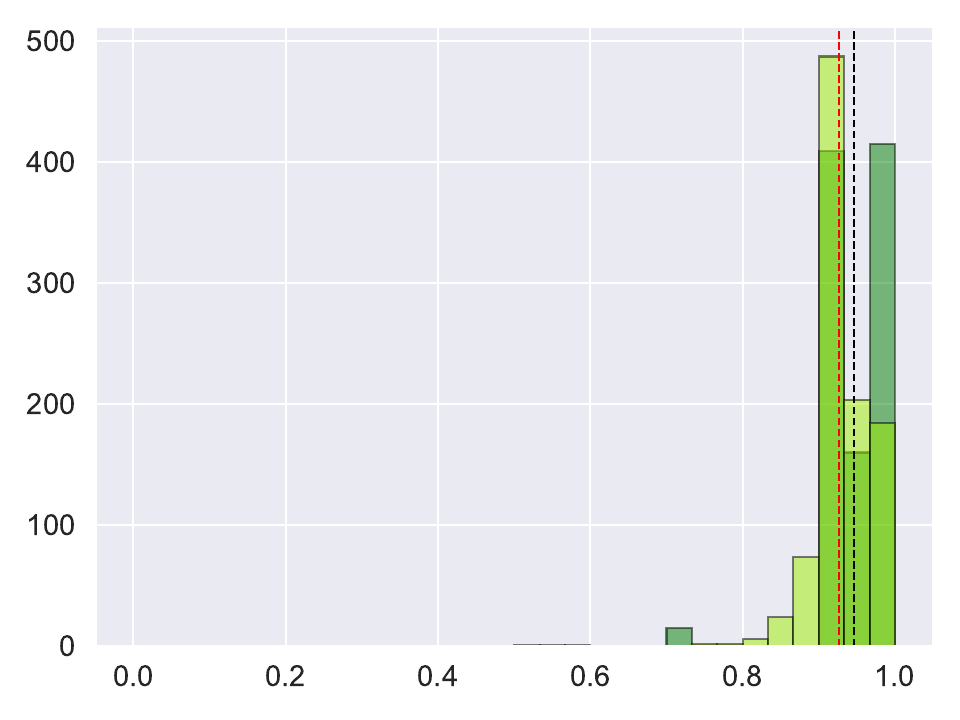}%
    }
    \subfigure[PC, DeepSeekCoder, \textbf{ECE: 0.1713}]{
    \includegraphics[scale=0.26]{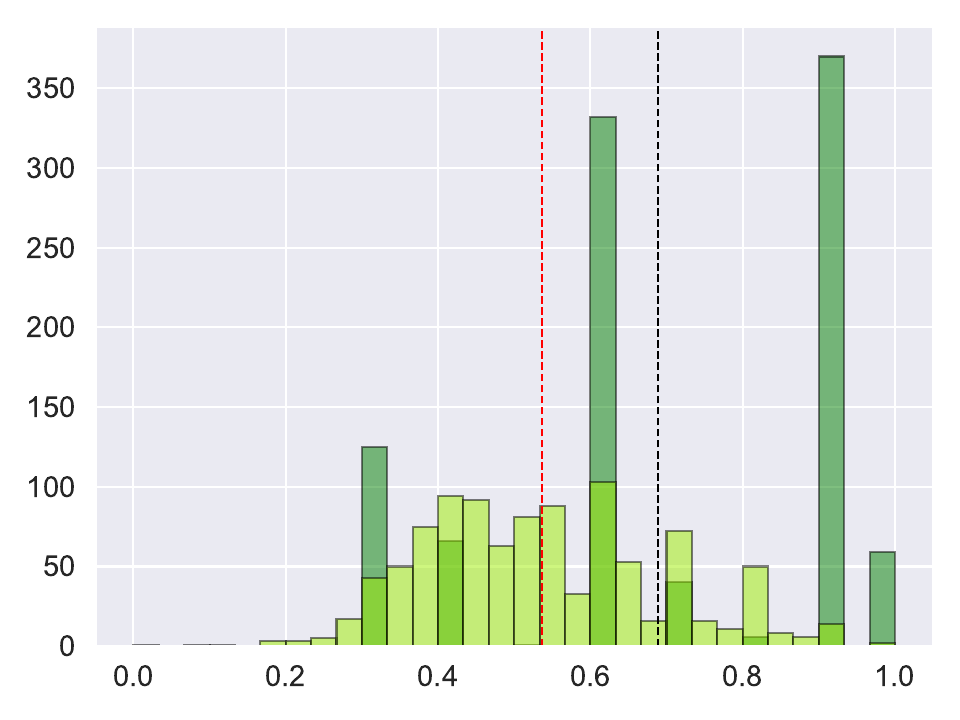}%
    }
    \subfigure[TMN, LLaMA3, \textbf{ECE: 0.0701}]{
    \includegraphics[scale=0.26]{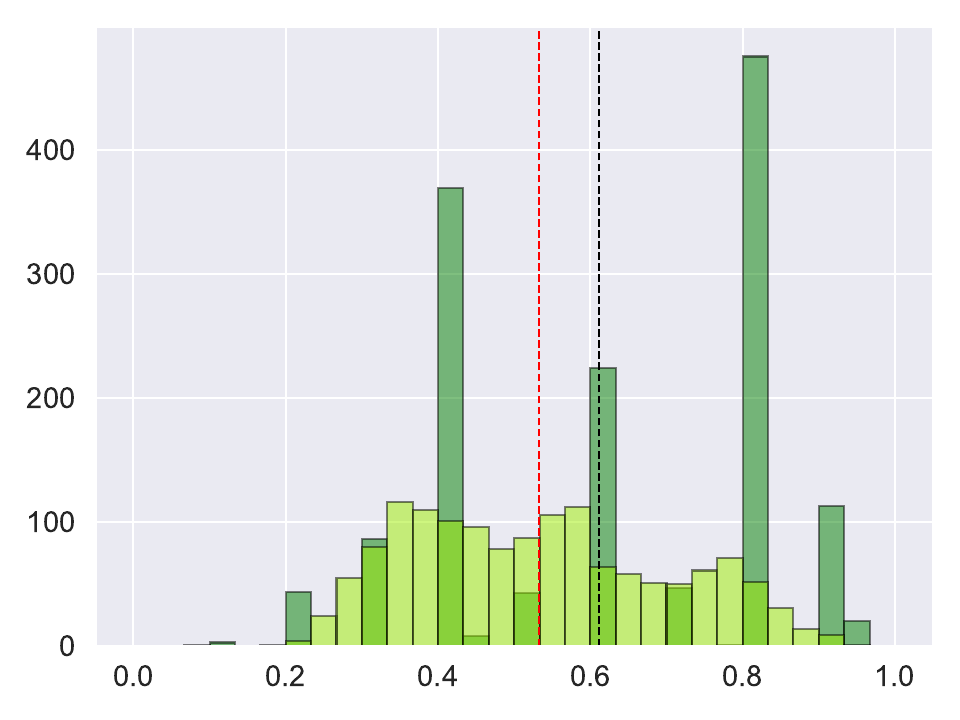}%
    }
    \subfigure[Sentiment, LLaMA3, \textbf{ECE: 0.2345}]{
    \includegraphics[scale=0.26]{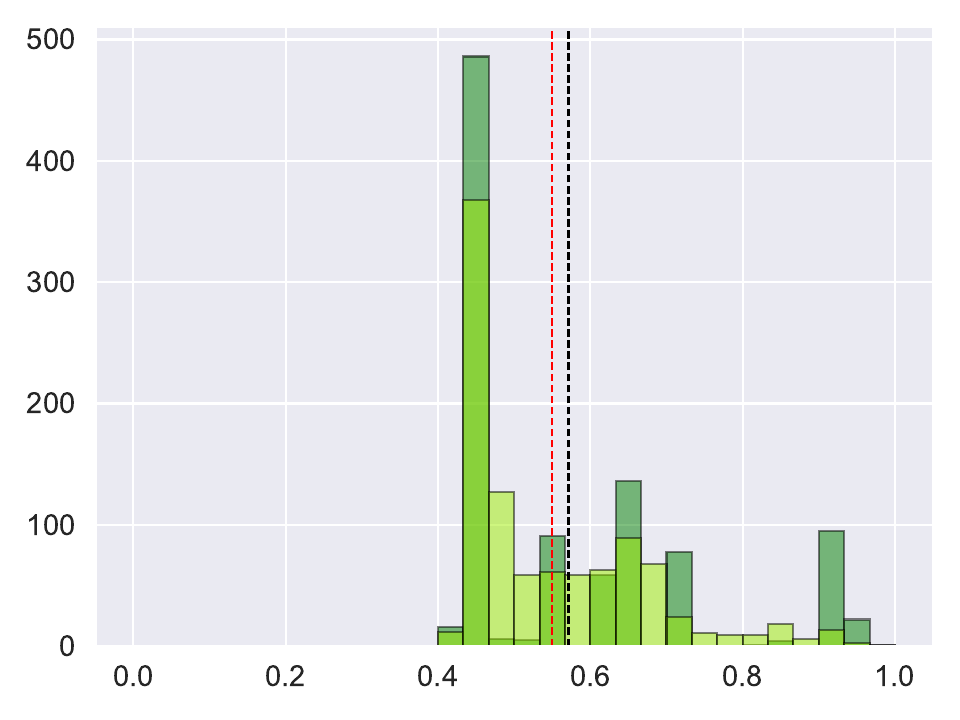}%
    }
    \subfigure[CD, GPT3.5, \textbf{ECE: 0.2563}]{
    \includegraphics[scale=0.26]{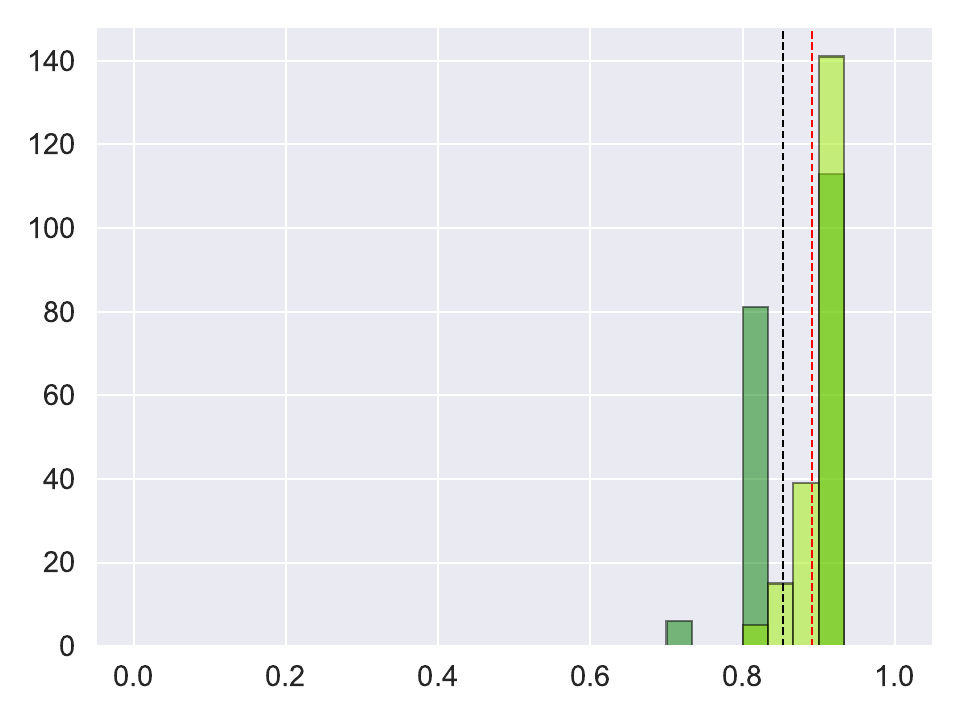}%
    }
    \subfigure[PC, GPT3.5, \textbf{ECE: 0.3449}]{
    \label{fig:rq2_confidence_pc_gpt35}
    \includegraphics[scale=0.26]{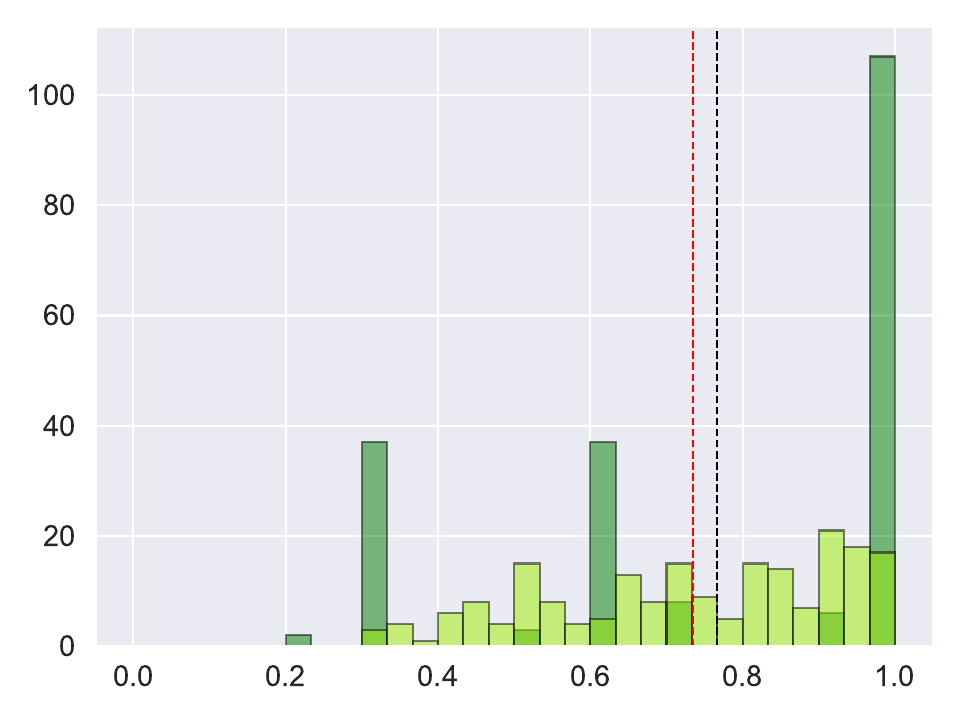}%
    }
    \subfigure[TMN, GPT3.5, \textbf{ECE: 0.1030}]{
    \includegraphics[scale=0.26]{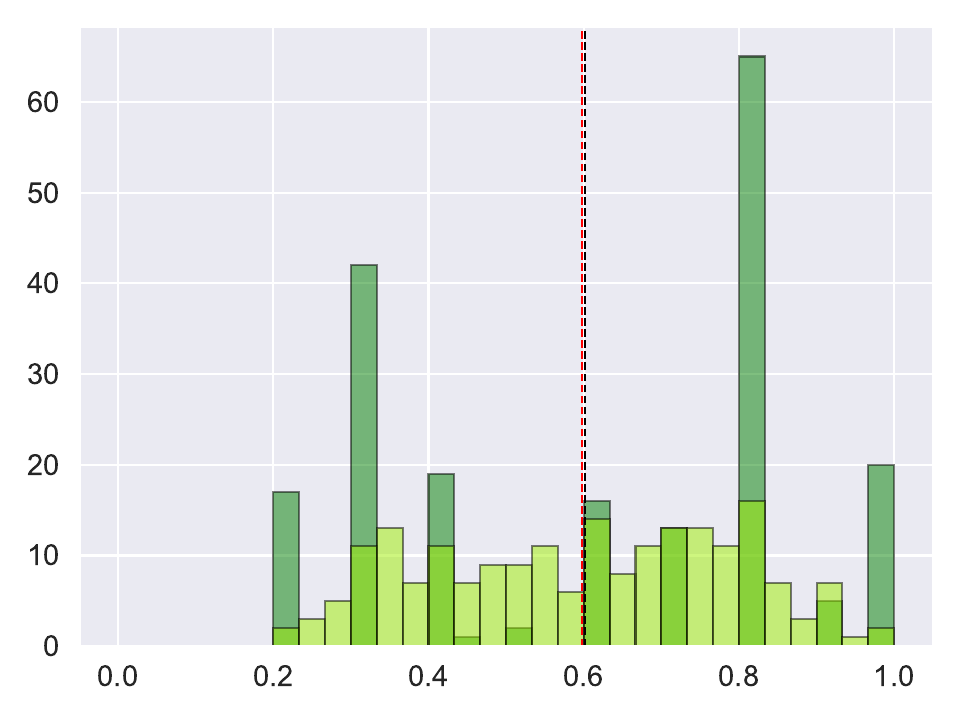}%
    }
    \subfigure[Sentiment, GPT3.5, \textbf{ECE: 0.1180}]{
    \includegraphics[scale=0.26]{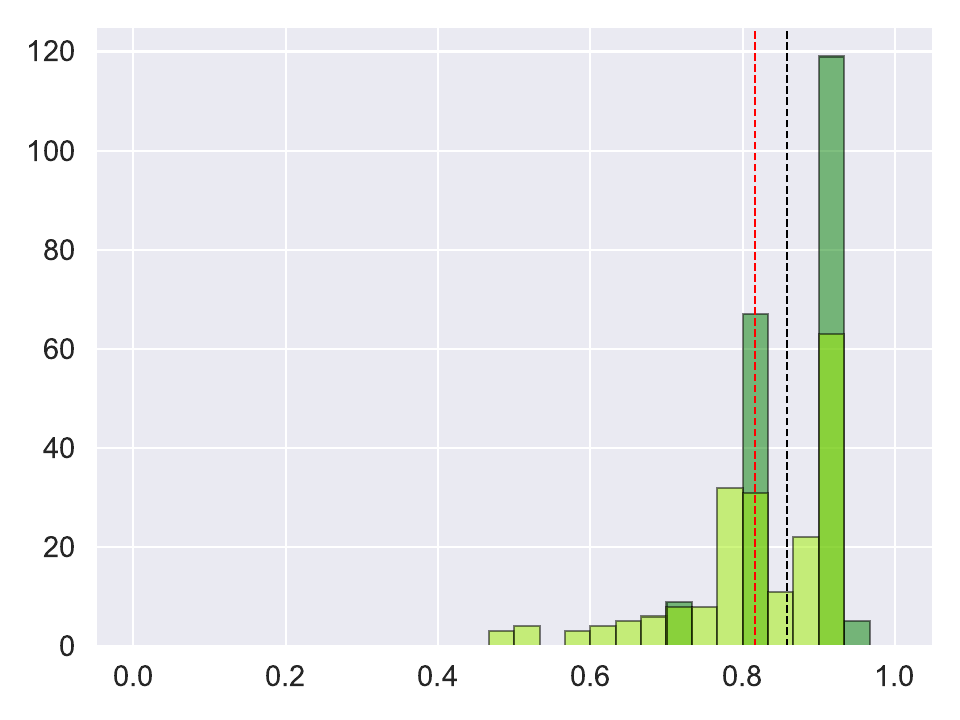}%
    }
    \subfigure[CD, GPT4, \textbf{ECE: 0.1638}]{
    \includegraphics[scale=0.26]{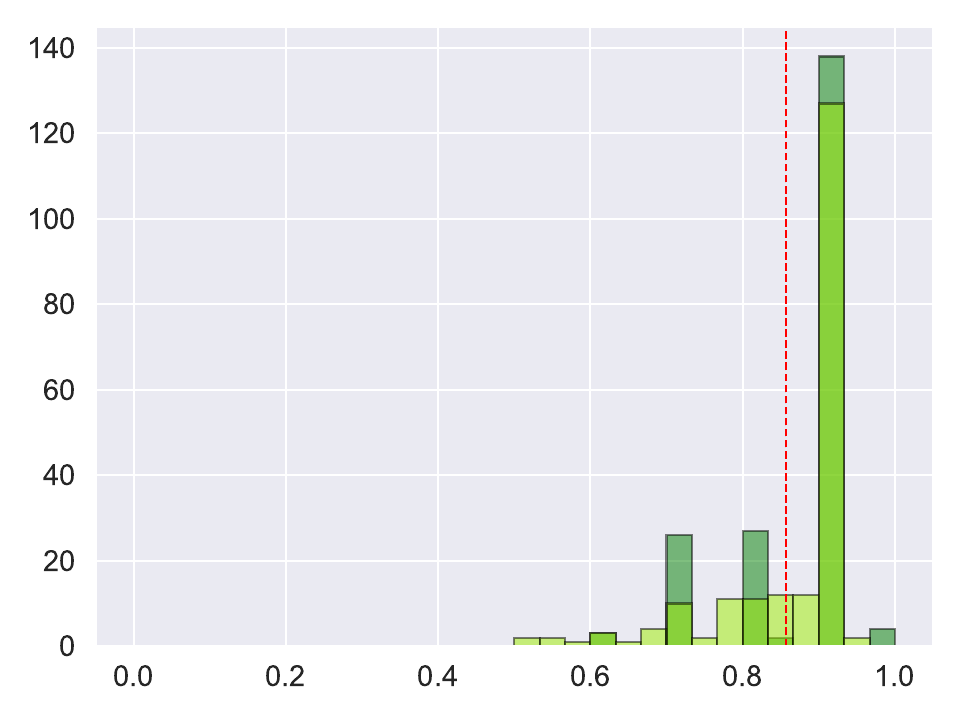}%
    }
    \subfigure[TMN, GPT4, \textbf{ECE: 0.0846}]{
    \includegraphics[scale=0.26]{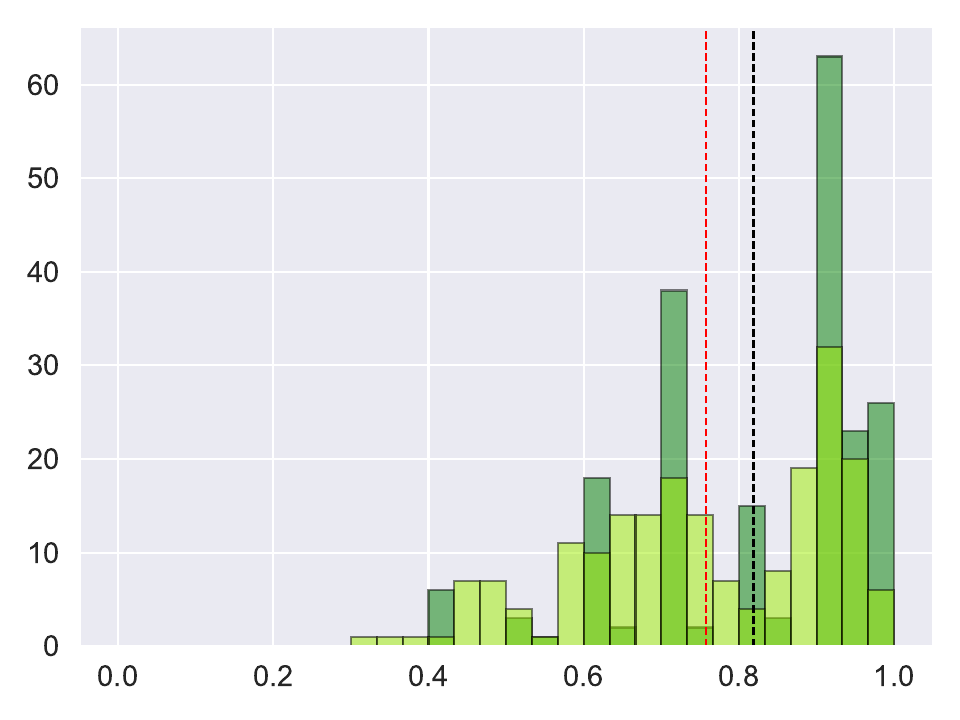}%
    }
    \subfigure[Sentiment, GPT4, \textbf{ECE: 0.1093}]{
    \includegraphics[scale=0.26]{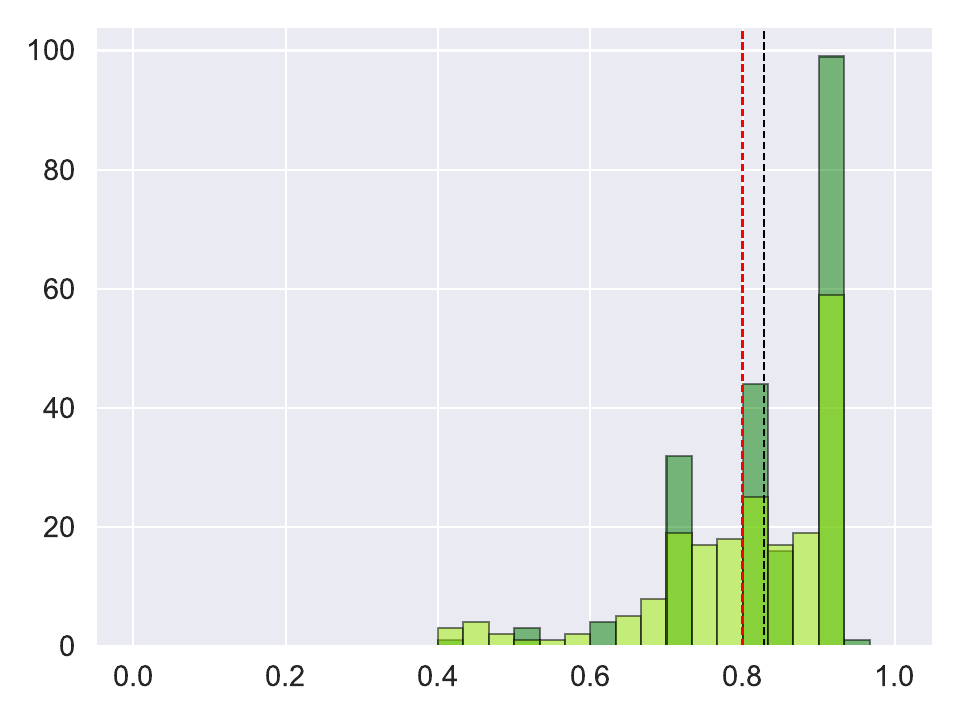}%
    }

    \caption{Distribution of prediction confidence and ECE for LLMs. \colorbox{mycolor_green}{Histogram}: confidence before smoothing.  \colorbox{mycolor}{Histogram}: confidence after smoothing. Black~(Red) dashed line: average confidence before~(after) smoothing.}
    \label{fig:rq2_confidence}
\end{figure*}

\begin{table*}[]
\centering
\caption{ECE and variance of prediction distribution before and after confidence smoothing.}
\label{table:rq2_confidence_diversity}
\resizebox{0.55\textwidth}{!}
{
\begin{tabular}{lcccc}
\bottomrule
\textbf{} & \multicolumn{2}{c}{\textbf{ECE ($\downarrow$)}} & \multicolumn{2}{c}{\textbf{Diversity ($\downarrow$)}} \\
\textbf{Model} & \textbf{Before} & \textbf{After} & \textbf{Before} & \textbf{After} \\ \bottomrule
CD-DeepSeekCoder & 0.9421 & 0.9112 & 11066.62 & 9535.22 \\
CD-GPT35 & 0.1915 & 0.2563 & 584.96 & 655.29 \\
CD-GPT4 & 0.1630 & 0.1638 & 696.62 & 542.42 \\
PC-DeepSeekCoder & 0.3110 & 0.1713 & 7963.02 & 1189.08 \\
PC-GPT35 & 0.4560 & 0.3449 & 205.36 & 24.69 \\
PC-GPT4 & 0.0143 & - & 332.00 & - \\
TMN-LLaMA3 & 0.1711 & 0.0701 & 12338.18 & 1641.37 \\
TMN-GPT35 & 0.0982 & 0.1030 & 205.36 & 24.69 \\
TMN-GPT4 & 0.0407 & 0.0846 & 196.56 & 62.29 \\
Sentiment-LLaMA3 & 0.5714 & 0.2345 & 8301.62 & 4898.62 \\
Sentiment-GPT35 & 0.0917 & 0.1180 & 335.87 & 104.87 \\
Sentiment-GP4 & 0.0787 & 0.1093 & 230.60 & 93.33 \\ \bottomrule
\textbf{Average change} & \multicolumn{2}{c}{\textbf{$\textcolor{green}{\uparrow}$ 17.60\%}} & \multicolumn{2}{c}{\textbf{$\textcolor{green}{\uparrow}$ 55.44\%}} \\ \bottomrule
\end{tabular}
}
\end{table*}

We further study how MuCS affects the prediction of LLMs and whether it diversifies the prediction confidence or not. Figure~\ref{fig:rq2_confidence} depicts the confidence of LLMs on the datasets before and after confidence smoothing and Table~\ref{table:rq2_confidence_diversity} summarizes the average ECE scores. The results show that even though MuCS can reduce the average ECE scores~(i.e., better calibration), there are still multiple pairs that have similar ECE scores, e.g., Sentiment-GPT35 and Sentiment-GPT4. Moreover, previous works~\cite{zhu2023rethinking} stated the effectiveness of fault detection methods is not directly related to the calibration of the model~(most confidence calibration methods are useless or harmful for fault detection). It is not clear whether the improvements are introduced by better calibration.

Then, we check if MuCS can increase the diversity of confidence distribution to solve the confidence concentrated issue. We compute the variance of the number of confidence scores in each range to check this diversity. The results presented in Table~\ref{table:rq2_confidence_diversity} demonstrated that, except for one case CD-GPT35, LLMs have more diverse predictions on the datasets with an average 55.44\% diversity improvement. Taking Figure~\ref{fig:rq2_confidence_pc_gpt35} as an example, the confidence scores of GPT3.5 are concentrated in 1.0 which is highly contradictory to 
its accuracy~(36\%). After the mutation-based smooth, confidence scores are spread throughout the distribution which is more useful to help understand the capability of LLMs for this task.

\noindent
\\
{\framebox{\parbox{0.96\linewidth}{
\textbf{Answer to RQ4}: MuCS significantly increases the diversity of confidence distribution of LLMs with an average 55.44\% improvement and eliminates the concentrated confidence problem. }}}


\section{Discussion}
\label{sec:discussion}

\subsection{Confidence vs. Correctness}

\begin{table}[]
\centering
\caption{Correlation between prediction confidence and prediction correctness.}
\label{table:rq4_correlation}
\resizebox{0.55\textwidth}{!}
{
\begin{tabular}{lcccc}
\bottomrule
 & \multicolumn{2}{c}{\textbf{Without MuCS}} & \multicolumn{2}{c}{\textbf{With MuCS}} \\
\textbf{Model} & \textbf{Significance} & \textbf{P-value} & \textbf{Significance} & \textbf{P-value} \\ \bottomrule
CD-DeepSeekCoder & 0.2753 & 7.43E-19 & 0.2243 & 1.09E-12 \\
CD-GPT3.5 & 0.2636 & 1.21E-13 & 0.3032 & 9.41E-18 \\
CD-GPT4 & 0.2969 & 4.71E-17 & 0.2823 & 1.70E-15 \\
PC-DeepSeekCoder & -0.0522 & 9.92E-02 & 0.2402 & 1.38E-14 \\
PC-GPT3.5 & 0.0952 & 5.22E-02 & 0.2546 & 1.07E-07 \\
TMN-LLaMA3 & 0.1110 & 2.53E-05 & 0.2426 & 1.21E-20 \\
TMN-GPT3.5 & 0.2904 & 2.97E-29 & 0.3238 & 2.47E-36 \\
TMN-GPT4 & 0.3909 & 1.55E-53 & 0.4059 & 5.97E-58 \\
Sentiment-LLaMA3 & 0.4603 & 1.34E-53 & 0.3730 & 2.33E-34 \\
Sentiment-GPT3.5 & 0.2007 & 4.37E-03 & 0.2124 & 2.53E-03 \\
Sentiment-GPT4 & 0.4263 & 3.10E-10 & 0.4332 & 1.49E-10 \\ \bottomrule
\end{tabular}
}
\end{table}

As mentioned before, all considered fault detection methods rely on the assumption -- there is a connection between the prediction confidence and the prediction correctness. However, it is unclear whether the assumption stands or not, especially for LLMs. We design an exploratory study to answer this problem. Specifically, we label the corrected prediction as 1 and the incorrect prediction as 0 and then utilize Pearson correlation coefficient to quantify the relationship between confidence and correctness as shown in Table~\ref{table:rq4_correlation}. We can see that there is a clear positive correlation between these two indicators except for the problem classification task before using MuCS. Moreover, after employing MuCS, all correlations become positive with a significance less than 0.03. This exploratory study demonstrates the relationship between confidence and correctness exists, and can be enhanced by MuCS.

\subsection{Limitation and Future Direction}
\label{sec:limitation}
\textbf{Limitation.} For a given input, our proposed mutation-based confidence smooth solution needs to access LLMs $n$ times, thus, the money consumption is $n$ times than the normal way when we test closed-source LLMs. However, from the view of executing time, the increased cost is negligible since it is possible to perform the prediction of a batch of data at the same time. There is indeed a trade-off between the monetary cost of accessing close-source LLMs and data labeling. This should be considered case by case.

\textbf{Future direction.} In this work, we only consider classification tasks since most existing fault detection methods are proposed for these tasks. Some methods, such as PRIMA~\cite{10.1109/ICSE43902.2021.00046} can be used for none classification tasks directly. Unfortunately, applying it to LLMs is not practical as it needs to mutate deep learning models but LLMs are much larger than classical deep learning models. However, the most important tasks people would use are generation tasks, e.g., chatting using ChatGPT. One of the important directions in this field is to propose fault detection methods for non-classification tasks, especially generation tasks, and then use these methods to help reveal critical faults in LLMs. Some metrics~\cite{kuhn2023semantic, huang2023look} have been designed for measuring the uncertainty in LLMs, how to combine these uncertainty metrics with existing methods could be the first try. 

\subsection{Threats to Validity}

The internal threat lies in the implementation of fault detection methods, LLMs, and prompt mutation methods. The implementation of each fault detection method comes from the official project provided in the original paper. The implementation of LLMs comes from the famous open-source project Hugging Face. The code of text mutation also comes from the commonly used project TextAugment~\cite{marivate2020improving}. The implementation of code refactoring methods modified from the previous work~\cite{dong2023mixcode}. 

The external threat comes from our studied tasks, datasets, LLMs, and fault detection methods. As mentioned in Section~\ref{sec:limitation}, we only consider the classification tasks in this work. For the tasks, we consider tasks from both NLP and SE fields. Two traditional NLP tasks and two programming tasks are included in our work. For the studied LLMs, we consider both open-source LLMs, e.g., LLaMA, and closed-source LLMs~(GPT4), where GPT4 is recognized as the SOTA LLM currently. Besides, another threat that comes from the datasets and models is that there might be data leakage issues in LLMs. For example, we found GPT4 has 100\% accuracy on the problem classification task which means our test set is likely in the training set of GPT4. However, it is difficult to check this issue as GPT3.5 and GPT4 are closed-source LLMs. We can only exclude the cases from the accuracy analysis of LLMs. For the fault detection methods, we follow the previous works~\cite{10.1145/3643678, ma2021test, hu2023evaluating} and explore 10 fault detection methods including both learning-based methods and output-based methods in our study. These methods are from different communities, for example, DeepGini is from the SE community and TestRank is from the ML community. We believe our studied objectives are representative and diverse enough and the conclusion drawn from this work can be generalized to other similar cases. 

The construct threat could be the hyperparameter settings and the non-determinism of LLMs. We directly use the default settings suggested by Hugging Face and OpenAI's API to build or access LLMs. Besides, we release the datasets and prompts used in our experiments to support reproducing results reported in our work and for future research.

\section{Conclusion}
\label{sec:conclusion}

In this work, we conducted the first empirical study to explore the potential of fault detection methods for LLMs. Based on the experiments on four tasks including NLP and code tasks, four LLMs, and nine fault detection methods, we found LLMs are overconfident with their predictions and simple methods such as Margin performs the best in revealing faults in LLMs. To enhance the detection performance, we proposed MuCS, which uses mutation analysis to smooth the prediction confidence of LLMs. Specifically, we utilized text transformation and code refactoring techniques to mutate the inputs and collect the averaged output probabilities of all mutants as the final prediction confidence of LLMs. Evaluation results demonstrated that MuCS significantly increased the diversity of output distribution of LLMs and enhanced the effectiveness of fault detection by up to 70.53\% in terms of test relative coverage score.

\bibliographystyle{ACM-Reference-Format}
\bibliography{main}

\end{document}